\documentclass[twocolumn]{aastex6}

\begin{document}

\title{A 2MASS/AllWISE Search for Extremely Red L Dwarfs -- The Discovery of Several Likely L Type Members of $\beta$ Pic, AB Dor, Tuc-Hor, Argus, and the Hyades}

\author{Adam C. Schneider\altaffilmark{1,2,3}, James Windsor\altaffilmark{4}, Michael C. Cushing\altaffilmark{4}, \& J. Davy Kirkpatrick\altaffilmark{5}}  

\altaffiltext{1}{School of Earth and Space Exploration, Arizona State University, Tempe, AZ, 85282, USA; aschneid10@gmail.com}
\altaffiltext{2}{Visiting Astronomer at the Infrared Telescope Facility, which is operated by the University of Hawaii under Cooperative Agreement no. NNX-08AE38A with the National Aeronautics and Space Administration, Science Mission Directorate, Planetary Astronomy Program.}
\altaffiltext{3}{Visiting astronomer, Cerro Tololo Inter-American Observatory, National Optical Astronomy Observatory, which is operated by the Association of Universities for Research in Astronomy (AURA) under a cooperative agreement with the National Science Foundation.}
\altaffiltext{4}{Department of Physics and Astronomy, University of Toledo, 2801 W. Bancroft St., Toledo, OH 43606, USA}
\altaffiltext{5}{IPAC, Mail Code 100-22, Caltech, 1200 E. California Blvd. Pasadena, CA 91125, USA}

\begin{abstract}

Young brown dwarfs share many properties with directly imaged giant extrasolar planets.  They therefore provide unique laboratories for investigating the full range of temperature and mass encompasses by the growing collection of planets discovered outside our Solar System.  Furthermore, if they can be tied to a particular group of coeval stars, they also provide vital anchor points for low-mass empirical isochrones.  We have developed a novel procedure for identifying such objects based on their unique 2MASS and AllWISE colors.  Using our search criteria, we have identified 50 new, late-type L dwarf candidates, 47 of which are spectroscopically confirmed as L dwarfs with follow-up near-infrared spectroscopy.  We evaluate the potential membership of these objects in nearby, young moving groups using their proper motions, photometric distance estimates, and spectroscopic indicators of youth, and find seven likely L-type members belonging to the $\beta$ Pictoris moving group, the AB Doradus moving group, the Tucana-Horologium association, or the Argus association, in addition to several lower probability members.  Also found are two late-type (L5 and L6) potential members of the nearby Hyades cluster (WISEA J043642.75$+$190134.8 and WISEA J044105.56$+$213001.5). 

\end{abstract}


\section{Introduction}

Recent studies have shown that young, late-type brown dwarfs can be used as proxies for young, giant extrasolar planets (e.g., \citealt{fah16}) because the same physics and chemistry governs the atmospheres of both sets of objects.  Free-floating brown dwarfs are typically much easier to observe than their exoplanetary counterparts because they do not compete with the bright glare of a nearby host star.  Furthermore, if a young brown dwarf can be tied to a nearby coeval moving group or cluster, it can serve as an evolutionary benchmark due to the fact that group properties, such as age and metallicity, can be applied to that object.  Low-mass members are also critical for constraining the low-mass end of the initial mass function (IMF) of coeval nearby groups (e.g., \citealt{gagne16}).  

Many of the young brown dwarfs in the literature have been found serendipitously through surveys for brown dwarfs or high proper motion objects (e.g., \citealt{kirk06}, \citealt{giz12}, \citealt{liu13}, \citealt{mace13}, \citealt{thomp13}, \citealt{schneid14}, \citealt{kell15}).  There are, however, recent efforts focused specifically on identifying young brown dwarfs (e.g., \citealt{gagne15a}, \citealt{aller16}).  These efforts typically use a combination of color criteria to select for late-type objects, and kinematic constraints associated with a particular association or associations to identify candidate moving group members.  Such searches have been adept at identifying late-M to early-L bona fide and candidate members of nearby, young moving groups.  These objects have ages ranging from $\sim$10 Myr to $\sim$150 Myr and, in some instances, have masses that extend into the planetary-mass regime (see Fig.\ 34 of \citealt{fah16}).   Despite these targeted searches, the latest spectral type members ($\geq$L5) remain particularly elusive.  The most likely reason for this is because such objects are much fainter than field age objects of the same spectral type, especially at the $J$ passband (see Figure 15 of \cite{liu16}).  However, it is exactly these late-type members that make the most ideal proxies as exoplanet analogs and dictate the shape of the IMF at the low mass end.  

One common trait among young, low-mass brown dwarfs is their red near- and mid-infrared colors compared to field age objects of the same spectral type (see e.g., Figures 5-14 in \citealt{fah16}).  These red colors are typically ascribed to enhanced amounts of dust and/or clouds in their atmospheres, effectively shifting their emergent flux to longer wavelengths.  The presence of excess clouds and dust is due to the lower surface gravities of these objects due to the fact that they are young and still contracting to their final radii.  Note that while red near-infrared colors are common among young brown dwarfs, there are examples of brown dwarfs with unusually red colors that are not believed to be young (e.g., \citealt{loop08}, \citealt{kirk10}, \citealt{mar14}), and for some objects, alternative explanations have been proposed (e.g., disk structures -- \citealt{zak17}).  Nevertheless, we sought to identify more young, nearby, late-type brown dwarfs based on their uniquely red colors.  We describe our search process in the following section, followed by a summary of our follow-up observations.  Lastly, we present the results of the survey and discuss individual objects of note.  

\section{Identifying Young Late-type L Dwarfs}
As noted in \cite{schneid14}, young, late-type ($>$L5) brown dwarfs occupy a unique region of Two Micron All Sky Survey (2MASS; \citealt{skrut06}) and {\it Wide-field Infrared Survey Explorer} ({\it WISE}; \citealt{wright10}) color space because they tend to have much redder colors than those of older field brown dwarfs, at least down to the L/T transition (\citealt{fah16}, \citealt{liu16}). We investigated this region of color space for additional candidate young, late-type brown dwarfs. Specifically, we searched for objects within the 2MASS and AllWISE point source catalogs with $J-K_{\rm S}$ $\geq$ 2.0 mag, $J-K_{\rm S}$ $\leq$ 3.5 mag, W1$-$W2 $\geq$ 0.3 mag, and W1$-$W2 $\leq$ 0.9 mag (see Figure 1).  This region was chosen to encompass known, red, young brown dwarfs, such as 2MASS J00470038$+$6803543 \citep{giz12}, 2MASS J03552337$+$1133437 \citep{reid06b}, WISE J174102.78$-$464225.5 \citep{schneid14}, and 2MASS J22443167$+$2043433 \citep{dahn02}, as well as the young companion VHS 1256$-$1257b \citep{gauza15}, while excluding the majority of known L dwarfs, as shown in Figure 1.  Note that we did not require objects to be detected in the $J$-band, anticipating the existence of objects so red in $J - K_{\rm S}$ color that they may only be detected at the $K_{\rm S}$-band\footnote{Note that in cases where $J$ band magnitudes are upper limits, such objects may have $J-K_{\rm S}$ colors redder than 3.5 mag.}.  We do, however, require that objects are well detected in the $K_S$, W1, and W2 bands (i.e., not upper limits and photometric uncertainties $\leq$0.2 mag).  We also avoid the galactic plane ($\vert b \vert$ $>$ 5.0 degrees) to exclude highly confused regions that would affect our 2MASS/AllWISE cross-match and star forming regions where the effects of reddening could be high, resulting in a significant increase of false-positives.  Lastly, we require that the separation between the 2MASS and AllWISE source positions be greater than 1$''$, thereby ensuring each candidate shows appreciable proper motion between the 2MASS and AllWISE epochs ($\gtrsim$100 mas yr$^{-1}$, considering the $\sim$10 year time baseline between 2MASS and AllWISE). While there are certainly moving group members with total proper motion magnitudes $<$100 mas yr$^{-1}$, we chose this limit so that the motion of our candidates could be clearly seen in our finder chart inspection process (see next paragraph) and to return a reasonable number of candidates for inspection.  These constraints returned 5,555 sources. 
  
\begin{figure*}
\plotone{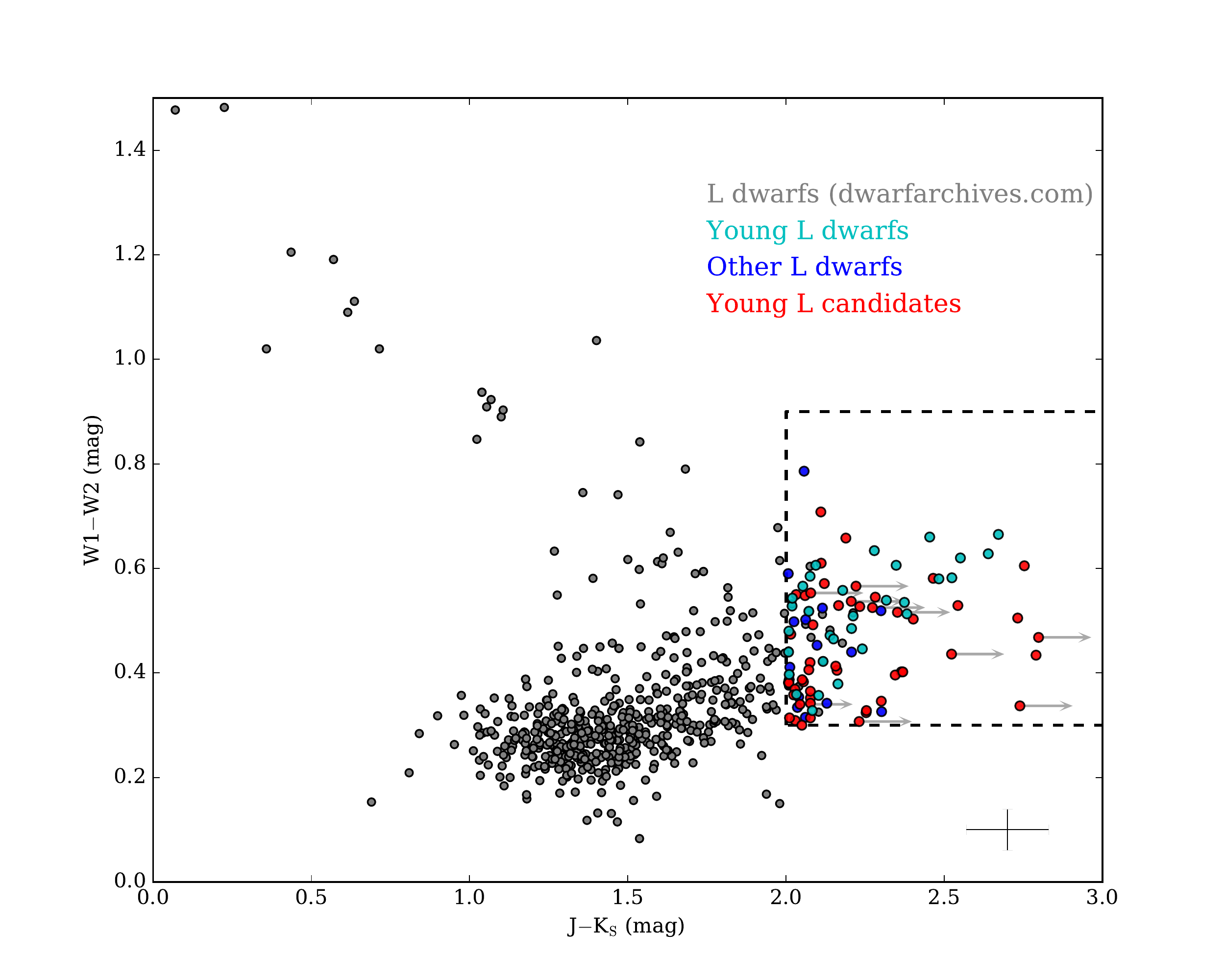}
\caption{The $J-K_{\rm s}$ versus W1$-$W2 color-color diagram showing our young L candidates (red) along with  recovered known L dwarfs from our search (young and otherwise -- light and dark blue, respectively), and all L dwarfs from dwarfarchives.com (grey). The dashed black line indicates the search region of this survey.  The typical photometric uncertainty for objects within our search region is plotted in the bottom right corner.}  
\end{figure*}

We then scrutinized each candidate individually by creating and inspecting a finder chart for each source using available optical Digitized Sky Survey (DSS), near-infrared (2MASS), and mid-infrared (AllWISE) images (see e.g., \citealt{schneid16a}) to ensure each candidate is a point source (i.e., not extended or blended) and has noticeable proper motion.  

We found 98 young brown dwarf candidates with this search.  Forty-eight of the candidates are previously known -- 47 are spectroscopically confirmed L dwarfs and one is WISEA J041847.95$+$252001.8, a highly reddened K dwarf \citep{kirk16}.  Table 1 lists these 48 known objects, their discovery references, and their spectral types.  Included in this list is the young L dwarf WISE 114724.10$-$204021.3 \citep{schneid16b}, the first discovery published from this survey.  Of the 47 known L dwarfs, 25 are either known to be young or show some signs of youth in their spectra.  This fraction (25/47, $\sim$53\%) is much higher than the young L dwarf fraction of $\sim$8\% reported by \cite{kirk08} in their analysis of field L dwarfs, showing the effectiveness of our method.  The 2MASS and AllWISE photometry for the remaining 50 candidates are listed in Table 2.  Note that some of these candidates have been identified as either high proper motion objects or brown dwarf candidates in the literature, but none have been spectroscopically confirmed.  Previous references to these objects are noted in Section 4.2.

\begin{deluxetable*}{lccccccc}
\tabletypesize{\footnotesize}
\tablecaption{Known Brown Dwarfs Recovered in this Survey}
\tablehead{
\colhead{AllWISE Designation} & \colhead{Other Name} & \colhead{Disc. Ref.} & \colhead{Sp.\ Type} & \colhead{Sp.\ Type Ref.} & \colhead{Young?\tablenotemark{a}}}
\startdata
J000627.85$+$185728.8 & \dots & \cite{schneid16a} & L7 & \cite{schneid16a} & N\\
J001851.52$+$515330.6 & PSO J004.7148$+$51.8918 & \cite{best15} &  L7 & \cite{best15} & Y?\\
J004701.09$+$680352.2 & 2MASS J00470038$+$6803543 & \cite{giz12} & L7 INT-G & \cite{giz15} & Y\\
J010332.31$+$193536.3 & 2MASSI J0103320$+$193536 & \cite{kirk00} & L6$\beta$ & \cite{fah12} & Y\\
J010752.84$+$004157.1 & 2MASS J01075242$+$0041563 & \cite{geb02} & L7 pec & \cite{gagne15b} & N\\
J012912.40$+$351757.3 & 2MASSW J0129122$+$351758 & \cite{kirk99} & L4 & \cite{kirk99} & Y?\\
J020503.72$+$125142.0 & 2MASSI J0205034$+$125142 & \cite{kirk00} & L6.5 & \cite{schneid14} & N\\
J020625.28$+$264023.5 & WISEPA J020625.26$+$264023.6 & \cite{kirk11} & L9 pec (red) & \cite{kirk11} & N?\\
J031854.39$-$342128.7 &  2MASS J03185403$-$3421292 & \cite{cruz07} & L7 & \cite{schneid14} & N\\
J032642.33$-$210207.3 & 2MASS J03264225$-$2102057 & \cite{giz03} & L5 $\beta$/$\gamma$ & \cite{gagne15b} & Y\\
J033703.75$-$175806.5 & 2MASS J03370359$-$1758079 & \cite{kirk00} & L4.5 & \cite{kirk00} & N\\  
J034909.45$+$151436.0 & PSO J057.2893$+$15.2433 & \cite{best15} &  L7 red & \cite{best15} & Y?\\
J035523.53$+$113337.2 & 2MASS J03552337$+$1133437 & \cite{reid06b} & L5$\gamma$ & \cite{fah13} & Y\\
J035822.61$-$411604.9 & 2MASS J03582255$-$4116060 & \cite{reid08} & L6 pec & \cite{gagne15b} & Y?\\ 
J040057.53$-$132204.2 & SIMP J04005763$-$1322024 & \cite{rob16} & L7:: & \cite{rob16} & N\\ 
J041847.95$+$252001.8 & \dots & \cite{kirk16} & $\sim$K (reddened) & \cite{kirk16} & N\\
J042107.45$-$630559.7 & 2MASS J04210718$-$6306022 & \cite{cruz07} & L5$\gamma$ & \cite{fah13} & Y\\
J044635.41$+$145125.8 & 2MASS J04463535$+$1451261 & \cite{hog08} & L2 & \cite{lod14} & Y\\
J050124.21$-$001047.1 & 2MASS J05012406$-$0010452 & \cite{reid08} & L3 VL-G & \cite{allers13} & Y\\
J074006.95$+$200920.0 & SDSS J074007.30$+$200921.9 & \cite{knapp04} & L6 & \cite{chiu06} & N\\   
J080958.86$+$443419.4 & SDSS J080959.01$+$443422.2 & \cite{knapp04} & L6 pec (red) & \cite{gagne15b} & Y?\\
J082029.81$+$450027.7 & 2MASS J08202996$+$4500315 & \cite{kirk00} & L7 & \cite{schneid14} & N\\
J082519.23$+$211548.3 & 2MASS J08251968$+$2115521 & \cite{kirk00} & L7 pec & \cite{gagne15b} & Y?\\
J082957.00$+$265509.2 & 2MASSW J0829570+265510 & \cite{kirk00} & L6.5 & \cite{kirk00} & Y?\\
J083542.14$-$081920.1 & 2MASS J08354256$-$0819237 & \cite{cruz03} & L4 pec & \cite{gagne15b} & Y?\\
J085757.94$+$570847.3 & 2MASS J08575849$+$5708514 & \cite{geb02} & L8 pec & \cite{gagne15b} & Y?\\
J095533.26$-$020841.6 & 2MASS J09553336$-$0208403 & \cite{gagne16} & L7 red & \cite{gagne16} & Y\\
J095608.17$-$144708.2 & PSO J149.0341$-$14.7857 & \cite{best15} & L9 & \cite{best15} & N\\
J095932.66$+$452329.3 & 2MASS J09593276$+$4523309 & \cite{zhang09} & L7.5 & \cite{zhang09} & Y\\
J100420.50$+$502257.6 & G 196$-$3B & \cite{reb98} & L2-L4$\gamma$ & \cite{gagne15b} & Y\\
J110233.55$-$235945.6 & 2MASS J11023375$-$2359464 & \cite{kirk00} & L4.5 & \cite{kirk00} & N\\
J111932.43$-$113747.7 & 2MASS J11193254$-$1137466 & \cite{kell15} & L7 red & \cite{kell16} & Y\\
J114724.10$-$204021.3 & \dots & \cite{schneid16b} & L7$\gamma$ & \cite{schneid16b} & Y\\
J125601.66$-$125728.7 & VHS 1256$-$1257b & \cite{gauza15} & L8 & \cite{gauza15} & Y\\
J130729.56$-$055815.4 & \dots & \cite{schneid16a} & L8 (sl.\ blue) & \cite{schneid16a} & N\\
J134316.31$+$394509.9 & 2MASSI J1343167$+$394508 & \cite{kirk00} & L5 & \cite{kirk00} & N\\
J155152.32$+$094114.2 & 2MASS J15515237$+$0941148 & \cite{reid08} & L4$\gamma$ & \cite{fah13} & Y\\
J155321.37$+$210908.4 & 2MASSW J1553214$+$210907 & \cite{kirk99} & L5.5 & \cite{kirk99} & N\\
J161542.44$+$495321.3 & 2MASS J16154255$+$4953211 & \cite{met08} & L3$-$L6$\gamma$ & \cite{gagne15b} & Y\\
J164715.57$+$563208.3 & WISEPA J164715.59$+$563208.2 & \cite{kirk11} & L9 pec (red) & \cite{kirk11} & N?\\
J172600.03$+$153818.2 & 2MASSI J1726000$+$153819 & \cite{kirk00} & L3$\beta$ & \cite{cruz09} & Y\\
J174102.77$-$464225.7 & WISE J174102.78$-$464225.5 & \cite{schneid14} & L5:$-$L7:$\gamma$ & \cite{gagne15b} & Y\\
J205202.06$-$204313.0 & \dots & \cite{schneid16a} & L8 (sl.\ blue) & \cite{schneid16a} & N\\
J214817.01$+$400404.1 & 2MASS J21481628$+$4003593 & \cite{loop08} & L6 (red) & \cite{schneid14} & N?\\
J215125.68$-$244100.5 & 2MASSI J2151254$-$244100 & \cite{lie06} & L4 pec & \cite{gagne15b} & Y?\\
J215434.68$-$105530.8 & SIMP J215434.5$-$105530.8 & \cite{gagne14b} & L5$\beta$/$\gamma$ & \cite{gagne15b} & Y\\
J224431.89$+$204340.2 & 2MASS J22443167$+$2043433 & \cite{dahn02} & L6$-$L8$\gamma$ & \cite{gagne15b} & Y\\
J234334.79$-$364603.4 & 2MASS J23433470$-$3646021 & \cite{gagne15b} & L3$-$L6$\gamma$ & \cite{gagne15b} & Y\\     
\enddata
\tablenotetext{a}{Previously determined signs of youth, where ``Y'' = young, ``Y?'' = likely young, ``N'' = not young, and, ``N?'' = unlikely to be young.}
\end{deluxetable*}

\begin{deluxetable*}{lccccccccc}
\tablecaption{2MASS and AllWISE Photometry of New Discoveries}
\tablehead{
\colhead{AllWISE Designation} &\colhead{$J$} & \colhead{$H$} & \colhead{$K_{\rm S}$} & \colhead{W1} & \colhead{W2}\\
 & (mag) & (mag) & (mag) & (mag) & (mag)}
\startdata
J002050.25$-$151913.1 & 16.962 $\pm$ 0.151 & 15.622 $\pm$ 0.102 & 14.933 $\pm$ 0.112 & 14.360 $\pm$ 0.030 & 14.051 $\pm$ 0.047 \\
J003052.08$-$380829.6 & 17.180 $\pm$ 0.231 & 16.062 $\pm$ 0.175 & 15.172 $\pm$ 0.162 & 14.898 $\pm$ 0.032 & 14.516 $\pm$ 0.051 \\
J004403.39$+$022810.6 & 16.997 $\pm$ 0.187 & 15.822 $\pm$ 0.169 & 14.876 $\pm$ 0.104 & 14.016 $\pm$ 0.027 & 13.445 $\pm$ 0.036 \\
J005811.69$-$565332.1 & 16.778 $\pm$ 0.165 & 15.554 $\pm$ 0.135 & 14.545 $\pm$ 0.094 & 13.763 $\pm$ 0.025 & 13.236 $\pm$ 0.028 \\
J010738.75$-$131413.7 & 16.710 $\pm$ 0.131 & 15.577 $\pm$ 0.120 & 14.625 $\pm$ 0.095 & 13.934 $\pm$ 0.027 & 13.442 $\pm$ 0.032 \\
J013556.99$-$620245.5 & 17.395 $\pm$ 0.271 & 16.187 $\pm$ 0.217 & 15.094 $\pm$ 0.138 & 14.793 $\pm$ 0.029 & 14.447 $\pm$ 0.041 \\
J014535.23$-$031412.9 & 17.124 $\pm$ 0.182 & 15.810 $\pm$ 0.140 & 14.958 $\pm$ 0.110 & 14.150 $\pm$ 0.027 & 13.621 $\pm$ 0.035 \\
J020047.29$-$510521.4 & 16.414 $\pm$ 0.124 & 14.941 $\pm$ 0.069 & 13.871 $\pm$ 0.052 & 12.885 $\pm$ 0.024 & 12.356 $\pm$ 0.023 \\
J020229.29$+$230513.9 & 17.221 $\pm$ 0.230 & 15.858 $\pm$ 0.129 & 15.206 $\pm$ 0.146 & 14.241 $\pm$ 0.027 & 13.767 $\pm$ 0.035 \\
J022609.16$-$161000.4 & 17.334 $\pm$ 0.266 & 15.750 $\pm$ 0.142 & 14.581 $\pm$ 0.093 & 13.745 $\pm$ 0.025 & 13.140 $\pm$ 0.027 \\
J023749.81$-$260543.8 & 16.777 $\pm$ 0.172 & 15.610 $\pm$ 0.135 & 14.768 $\pm$ 0.121 & 14.229 $\pm$ 0.026 & 13.848 $\pm$ 0.032 \\
J025954.88$-$314655.6 & 17.487 $\pm$ 0.239 & 16.380 $\pm$ 0.184 & 15.437 $\pm$ 0.169 & 14.919 $\pm$ 0.030 & 14.619 $\pm$ 0.047 \\
J032049.31$-$532656.7 & 17.032 $\pm$ 0.225 & 15.664 $\pm$ 0.128 & 14.871 $\pm$ 0.129 & 14.162 $\pm$ 0.026 & 13.757 $\pm$ 0.029 \\
J032440.23$-$191905.6 & 17.007 $\pm$ 0.205 & 15.591 $\pm$ 0.154 & 14.605 $\pm$ 0.111 & 14.125 $\pm$ 0.027 & 13.622 $\pm$ 0.031 \\
J041232.77$+$104408.3 & 17.606 $\pm$ 0.255 & 16.144 $\pm$ 0.164 & 15.242 $\pm$ 0.113 & 14.270 $\pm$ 0.029 & 13.868 $\pm$ 0.040 \\
J042231.34$+$081012.7 & 17.272 $\pm$ 0.226 & $>$16.043 & 15.240 $\pm$ 0.135 & 15.618 $\pm$ 0.050 & 15.068 $\pm$ 0.095 \\
J042506.66$-$425509.6 & 16.616 $\pm$ 0.135 & 15.011 $\pm$ 0.060 & 14.427 $\pm$ 0.067 & 13.477 $\pm$ 0.024 & 12.819 $\pm$ 0.024 \\
J043642.75$+$190134.8 & 17.121 $\pm$ 0.175 & 15.657 $\pm$ 0.098 & 14.868 $\pm$ 0.094 & 14.193 $\pm$ 0.030 & 13.869 $\pm$ 0.043 \\
J043718.77$-$550944.0 & 16.985 $\pm$ 0.192 & 15.583 $\pm$ 0.157 & 14.640 $\pm$ 0.098 & 14.135 $\pm$ 0.024 & 13.739 $\pm$ 0.026 \\
J044105.56$+$213001.5 & 17.274 $\pm$ 0.218 & 16.141 $\pm$ 0.168 & 15.197 $\pm$ 0.130 & 14.554 $\pm$ 0.032 & 14.202 $\pm$ 0.052 \\
J045900.42$-$285338.3 & 17.429 $\pm$ 0.282 & 16.375 $\pm$ 0.249 & 15.318 $\pm$ 0.197 & 14.305 $\pm$ 0.026 & 13.695 $\pm$ 0.028 \\
J050259.73$-$610206.1 & 17.087 $\pm$ 0.187 & 15.945 $\pm$ 0.149 & 15.010 $\pm$ 0.151 & 14.698 $\pm$ 0.026 & 14.356 $\pm$ 0.033 \\
J055959.30$-$583546.0 & 16.686 $\pm$ 0.144 & 15.416 $\pm$ 0.097 & 14.631 $\pm$ 0.090 & 14.450 $\pm$ 0.026 & 14.067 $\pm$ 0.032 \\
J065935.80$+$771457.8 & 16.865 $\pm$ 0.171 & 15.594 $\pm$ 0.115 & 14.708 $\pm$ 0.096 & 14.215 $\pm$ 0.026 & 13.802 $\pm$ 0.033 \\
J070534.00$-$183925.6 & 16.778 $\pm$ 0.129 & 15.498 $\pm$ 0.092 & 14.701 $\pm$ 0.092 & 13.977 $\pm$ 0.028 & 13.612 $\pm$ 0.035 \\
J071138.88$+$370601.0 & 17.165 $\pm$ 0.243 & 15.466 $\pm$ 0.131 & 14.911 $\pm$ 0.094 & 14.414 $\pm$ 0.030 & 14.086 $\pm$ 0.053 \\
J072352.62$-$330943.5 & 15.743 $\pm$ 0.059 & 14.471 $\pm$ 0.043 & 13.715 $\pm$ 0.047 & 13.068 $\pm$ 0.023 & 12.699 $\pm$ 0.023 \\
J081322.19$-$152203.2 &  $>$17.658 & 16.252 $\pm$ 0.183 & 14.860 $\pm$ 0.126 & 13.984 $\pm$ 0.026 & 13.516 $\pm$ 0.030 \\
J082624.09$-$601202.8 & $>$17.559 & 16.133 $\pm$ 0.205 & 14.820 $\pm$ 0.126 & 14.517 $\pm$ 0.026 & 14.180 $\pm$ 0.033 \\
J090258.99$+$670833.1 & 16.979 $\pm$ 0.246 & 15.089 $\pm$ 0.106 & 14.247 $\pm$ 0.108 & 13.200 $\pm$ 0.025 & 12.695 $\pm$ 0.026 \\
J093858.10$+$761211.5 & 16.984 $\pm$ 0.181 & 15.595 $\pm$ 0.124 & 14.908 $\pm$ 0.106 & 14.275 $\pm$ 0.026 & 13.855 $\pm$ 0.034 \\
J120104.57$+$573004.2 & 17.355 $\pm$ 0.235 & 16.407 $\pm$ 0.266 & 15.245 $\pm$ 0.133 & 14.368 $\pm$ 0.028 & 13.660 $\pm$ 0.032 \\
J130523.06$-$395104.9 & $>$17.107 & 16.138 $\pm$ 0.224 & 15.029 $\pm$ 0.133 & 14.193 $\pm$ 0.027 & 13.640 $\pm$ 0.032 \\
J131845.58$+$362614.0 & 17.212 $\pm$ 0.209 & 15.844 $\pm$ 0.136 & 15.161 $\pm$ 0.117 & 14.538 $\pm$ 0.028 & 14.151 $\pm$ 0.041 \\
J143211.17$+$324433.8 & $>$17.359 & 16.081 $\pm$ 0.186 & 15.138 $\pm$ 0.126 & 14.300 $\pm$ 0.027 & 13.734 $\pm$ 0.030 \\
J145642.68$+$645009.7 & 17.564 $\pm$ 0.310 & 15.631 $\pm$ 0.115 & 14.774 $\pm$ 0.102 & 13.888 $\pm$ 0.025 & 13.454 $\pm$ 0.027 \\
J153358.52$+$475706.9 & $>$17.280 & 15.909 $\pm$ 0.152 & 14.928 $\pm$ 0.152 & 14.374 $\pm$ 0.025 & 13.858 $\pm$ 0.031 \\
J162341.27$-$740230.4 & $>$17.075 & 15.481 $\pm$ 0.147 & 14.869 $\pm$ 0.131 & 13.923 $\pm$ 0.027 & 13.386 $\pm$ 0.030 \\
J173453.90$-$481357.9 & 16.285 $\pm$ 0.127 & 14.865 $\pm$ 0.075 & 13.916 $\pm$ 0.052 & 12.968 $\pm$ 0.025 & 12.566 $\pm$ 0.026 \\
J174057.82$+$131709.4 & $>$17.468 & 16.257 $\pm$ 0.212 & 15.195 $\pm$ 0.149 & 14.764 $\pm$ 0.033 & 14.239 $\pm$ 0.046 \\
J190722.56$+$472745.3 & 16.402 $\pm$ 0.124 & 15.269 $\pm$ 0.110 & 14.330 $\pm$ 0.073 & 13.862 $\pm$ 0.024 & 13.456 $\pm$ 0.027 \\
J201204.11$+$672608.0 & 17.148 $\pm$ 0.239 & 15.850 $\pm$ 0.179 & 15.124 $\pm$ 0.137 & 14.336 $\pm$ 0.025 & 13.978 $\pm$ 0.030 \\
J201530.67$-$421542.5 & 17.648 $\pm$ 0.317 & 16.233 $\pm$ 0.151 & 15.366 $\pm$ 0.145 & 14.592 $\pm$ 0.030 & 14.047 $\pm$ 0.041 \\
J201826.00$-$332207.3 & $>$17.33 & 15.908 $\pm$ 0.210 & 15.286 $\pm$ 0.147 & 15.153 $\pm$ 0.038 & 14.813 $\pm$ 0.073 \\
J204902.80$-$745613.5 & $>$17.789 & 16.328 $\pm$ 0.229 & 15.266 $\pm$ 0.163 & 14.595 $\pm$ 0.028 & 14.159 $\pm$ 0.035 \\
J225333.00$-$253948.0 & 17.152 $\pm$ 0.214 & 15.752 $\pm$ 0.148 & 15.075 $\pm$ 0.144 & 14.544 $\pm$ 0.029 & 14.230 $\pm$ 0.047 \\
J232307.08$+$054113.0 & 17.600 $\pm$ 0.276 & 15.961 $\pm$ 0.165 & 15.540 $\pm$ 0.173 & 14.642 $\pm$ 0.032 & 14.094 $\pm$ 0.043 \\
J232453.73$+$503525.4 & $>$17.084 & 15.868 $\pm$ 0.181 & 14.853 $\pm$ 0.104 & 14.449 $\pm$ 0.027 & 14.142 $\pm$ 0.035 \\
J233333.46$+$025128.4 & 16.688 $\pm$ 0.127 & 15.407 $\pm$ 0.093 & 14.677 $\pm$ 0.086 & 14.158 $\pm$ 0.028 & 13.844 $\pm$ 0.042 \\
J235422.31$-$081129.7 & 17.255 $\pm$ 0.230 & 15.962 $\pm$ 0.150 & 14.790 $\pm$ 0.119 & 13.962 $\pm$ 0.027 & 13.381 $\pm$ 0.033 \\
\enddata
\end{deluxetable*}

\section{Observations}

\subsection{Infrared Telescope Facility (IRTF)/SpeX}

Twenty-seven objects were observed using the SpeX spectrograph \citep{rayner03} at NASA's 3m IRTF.  All observations were made in prism mode with a 0\farcs5-wide slit, resulting in a resolving power ($\lambda$/$\Delta$$\lambda$) of $\sim$150 over the 0.8$-$2.5 $\mu$m range.  For each observation, the slit was oriented along the parallactic angle and exposures were taken at two different nod positions.  A0V stars at similar airmasses were observed immediately after each science target for telluric correction purposes.  Each spectrum was reduced using the SpeXtool reduction package (\citealt{vacca03}; \citealt{cush04}).  A summary of all IRTF/SpeX observations, including observation dates and exposure times, is given in Table 3.

\subsection{Magellan/FIRE}

Nine objects were observed with the Folded-port Infrared Echellette (FIRE; \citealt{sim13}) spectrograph located at the 6.5m Baade Magellan telescope.  All observations were made with the high-throughput prism mode, which achieved a resolving power of $\sim$450 across the 0.8$-$2.45 $\mu$m range.  We used the 0\farcs6 slit, aligned to the parallactic angle, and took exposures at two different nod positions along the slit.  For all science targets, we used the sample-up-the-ramp mode.  A0V stars were observed after each science target to correct for telluric absorption.  All reductions were performed using a modified version of the SpeXtool reduction package (\citealt{vacca03}; \citealt{cush04}).  A summary of all Magellan/FIRE observations, including observation dates and exposure times, is given in Table 3.
 
\subsection{CTIO Blanco 4m/ARCoIRIS}

Twenty-two objects were observed with the Astronomy Research using the Cornell Infrared Imaging Spectrograph (ARCoIRIS) on the 4m Blanco telescope located at the Cerro Tololo Inter-American Observatory (CTIO).  ARCoIRIS takes simultaneous spectra across 6 cross-dispersed orders covering the 0.8$-$2.4 $\mu$m range, with a resolving power of $\sim$3500.  Science exposures were taken at two different nod positions along the slit, which has a fixed width of 1\farcs1. After each science target, A0V stars were observed in order to execute telluric corrections.  Reductions were performed using a modified version of the SpeXtool reduction package (\citealt{vacca03}; \citealt{cush04}).  A summary of all ARCoIRIS observations, including observation dates and exposure times, is given in Table 3.
   
\begin{deluxetable*}{lcccc}
\tablecaption{Summary of Observations}
\tablehead{
\colhead{AllWISE Designation} & \colhead{Telescope/} & \colhead{Obs. Date} & \colhead{Exp. Time} &\colhead{S/N$_{\rm J}$\tablenotemark{a}}\\
 &  Instrument &  (UT) & (s) & }
\startdata
J002050.25$-$151913.1 & CTIO/ARCoIRIS & 2016 December 10 & 1440 & 3\\
J003052.08$-$380829.6 & Magellan/FIRE & 2016 July 18 & 1014 & 29\\
J003052.08$-$380829.6 & CTIO/ARCoIRIS & 2016 August 22 & 2160 & 11\\
J004403.39$+$022810.6 & IRTF/SpeX & 2016 September 29 & 2160 & 20\\
J005811.69$-$565332.1 & Magellan/FIRE & 2016 July 17 & 1014 & 26\\
J005811.69$-$565332.1 & CTIO/ARCoIRIS & 2016 August 19 & 2160 & 5\\
J010738.75$-$131413.7 & IRTF/SpeX & 2016 September 30 & 2640 & 6\\
J013556.99$-$620245.5 & Magellan/FIRE & 2016 July 17 & 1014 & 19\\
J013556.99$-$620245.5 & CTIO/ARCoIRIS & 2016 August 22 & 2160 & 8\\
J014535.23$-$031412.9 & IRTF/SpeX & 2016 September 29 & 1440 & 19\\
J020047.29$-$510521.4 & Magellan/FIRE & 2016 July 17 & 761 & 55\\
J020047.29$-$510521.4 & CTIO/ARCoIRIS & 2016 August 22 & 960 & 15\\
J020229.29$+$230513.9 & IRTF/SpeX & 2016 September 28 & 2160 & 24\\
J022609.16$-$161000.4 & Magellan/FIRE & 2016 July 18 & 1014 & 22\\
J022609.16$-$161000.4 & IRTF/SpeX & 2016 September 29 & 2160 & 17\\
J023749.81$-$260543.8 & CTIO/ARCoIRIS & 2016 December 11 & 2160 & 10\\
J025954.88$-$314655.6 & CTIO/ARCoIRIS & 2016 December 10 & 2160 & 8\\
J032049.31$-$532656.7 & CTIO/ARCoIRIS & 2016 December 11 & 2160 & 11\\
J032440.23$-$191905.6 & IRTF/SpeX & 2016 September 28 & 2160 & 36\\
J041232.77$+$104408.3 & IRTF/SpeX & 2016 September 30 & 2880 & 12\\
J042231.34$+$081012.7 & IRTF/SpeX & 2016 September 29 & 2160 & 17\\
J042506.66$-$425509.6 & CTIO/ARCoIRIS & 2016 January 25 & 2040 & 12\\
J043642.75$+$190134.8 & IRTF/SpeX & 2016 February 12 & 1440 & 12\\
J043718.77$-$550944.0 & CTIO/ARCoIRIS & 2016 December 11 & 2160 & 10\\
J044105.56$+$213001.5 & IRTF/SpeX & 2016 September 28 & 2880 & 10\\
J045900.42$-$285338.3 & CTIO/ARCoIRIS & 2016 December 10 & 2160 & 4\\
J050259.73$-$610206.1 & CTIO/ARCoIRIS & 2016 December 9 & 2160 & 5\\
J055959.30$-$583546.0 & CTIO/ARCoIRIS & 2016 December 10 & 2160 & 11\\
J070534.00$-$183925.6 & IRTF/SpeX & 2016 February 12 & 1440 & 18\\
J071138.88$+$370601.0 & IRTF/SpeX & 2016 February 12 & 1440 & 22\\
J072352.62$-$330943.5 & IRTF/SpeX & 2016 February 12 & 1440 & 53\\
J081322.19$-$152203.2 & CTIO/ARCoIRIS & 2016 December 11 & 2160 & 8\\
J082624.09$-$601202.8 & CTIO/ARCoIRIS & 2016 December 11 & 2160 & 10\\
J090258.99$+$670833.1 & IRTF/SpeX & 2016 February 12 & 2160 & 56\\
J095533.26$-$020841.6 & IRTF/SpeX & 2016 February 12 & 2160 & 17\\
J120104.57$+$573004.2 & IRTF/SpeX & 2016 February 12 & 2160 & 22\\
J130523.06$-$395104.9 & IRTF/SpeX & 2016 February 12 & 1440 & 7\\
J131845.58$+$362614.0 & IRTF/SpeX & 2016 February 12 & 1440 & 30\\
J145642.68$+$645009.7 & IRTF/SpeX & 2016 February 12 & 1920 & 31\\
J143211.17$+$324433.8 & IRTF/SpeX & 2016 May 31 & 2160 & 16\\
J153358.52$+$475706.9 & IRTF/SpeX & 2016 May 31 & 2160 & 16\\
J162341.27$-$740230.4 & Magellan/FIRE & 2016 July 17 & 1268 & 42\\
J162341.27$-$740230.4 & CTIO/ARCoIRIS & 2016 August 22 & 2160 & 6\\
J173453.90$-$481357.9 & CTIO/ARCoIRIS & 2016 August 19 & 1440 & 10\\
J174057.82$+$131709.4 & IRTF/SpeX & 2016 May 31 & 1440 & 17\\
J190722.56$+$472745.3 & IRTF/SpeX & 2016 June 26 & 2160 & 39\\
J201204.11$+$672608.0 & IRTF/SpeX & 2016 June 26 & 2160 & 8\\
J201530.67$-$421542.5 & CTIO/ARCoIRIS & 2016 June 21 & 1440 & 8\\
J201826.00$-$332207.3 & Magellan/FIRE & 2016 July 17 & 1268 & 34\\
J201826.00$-$332207.3 & CTIO/ARCoIRIS & 2016 August 22 & 2160 & 10\\
J204902.80$-$745613.5 & CTIO/ARCoIRIS & 2016 June 22 & 2880 & 6\\
J225333.00$-$253948.0 & Magellan/FIRE & 2016 July 18 & 1014 & 38\\
J225333.00$-$253948.0 & CTIO/ARCoIRIS & 2016 August 19 & 2160 & 5\\
J232307.08$+$054113.0 & Magellan/FIRE & 2016 July 18 & 1014 & 11\\
J232307.08$+$054113.0 & IRTF/SpeX & 2016 September 29 & 2160 & 19\\
J233333.46$+$025128.4 & IRTF/SpeX & 2016 September 28 & 1440 & 29\\
J235422.31$-$081129.7 & IRTF/SpeX & 2016 September 29 & 2160 & 21\\
J235422.31$-$081129.7 & CTIO/ARCoIRIS & 2016 August 22 & 2160 & 7\\
\enddata
\tablenotetext{a}{Signal-to-noise ratio achieved between 1.27 and 1.29 $\mu$m.}
\end{deluxetable*}

\section{Analysis}

\subsection{Basic Properties of the Entire Sample}

A total of 50 L-type brown dwarf candidates were found with this survey, 47 of which we have observed spectroscopically and confirm as L dwarfs. We determine spectral types for all near-infrared spectra following the method outlined in the Appendix of \cite{schneid14}, whereby we compare each spectrum via a $\chi$$^2$ fit to every near-infrared L dwarf spectral standard from \cite{kirk10} normalized between 1.27 and 1.29 $\mu$m.  Each spectrum is then inspected by-eye to find the best fit, with the results of the $\chi$$^2$ fitting as a guide.  The uncertainties after the by-eye inspection are $\pm$0.5 subtypes, except in cases of low signal-to-noise (S/N$<$10), where we assume a $\pm$1 subtype uncertainty.  All spectral types are given in Table 4 and comparisons of each acquired spectrum with the corresponding near-infrared spectral standards from the Spex Prism Spectral Library\footnote{http://pono.ucsd.edu/$\sim$adam/browndwarfs/spexprism/library.html} are shown in Figures 2$-$5.  If more than one spectrum was obtained for an object, the spectrum shown is that with the higher S/N.  For the 10 objects for which we have multiple spectra, all spectral types were consistent between observations.  Any ARCoIRIS spectrum shown has been smoothed to a similar resolution as the near-infrared spectral standards for comparison.  

We calculate the proper motion of each object using the positions from the AllWISE and 2MASS source catalogs.  Proper motion uncertainties are determined from the positional uncertainties for each object provided in the AllWISE and 2MASS catalogs.  Proper motions and uncertainties are given in Table 4.     

\begin{figure*}
\plotone{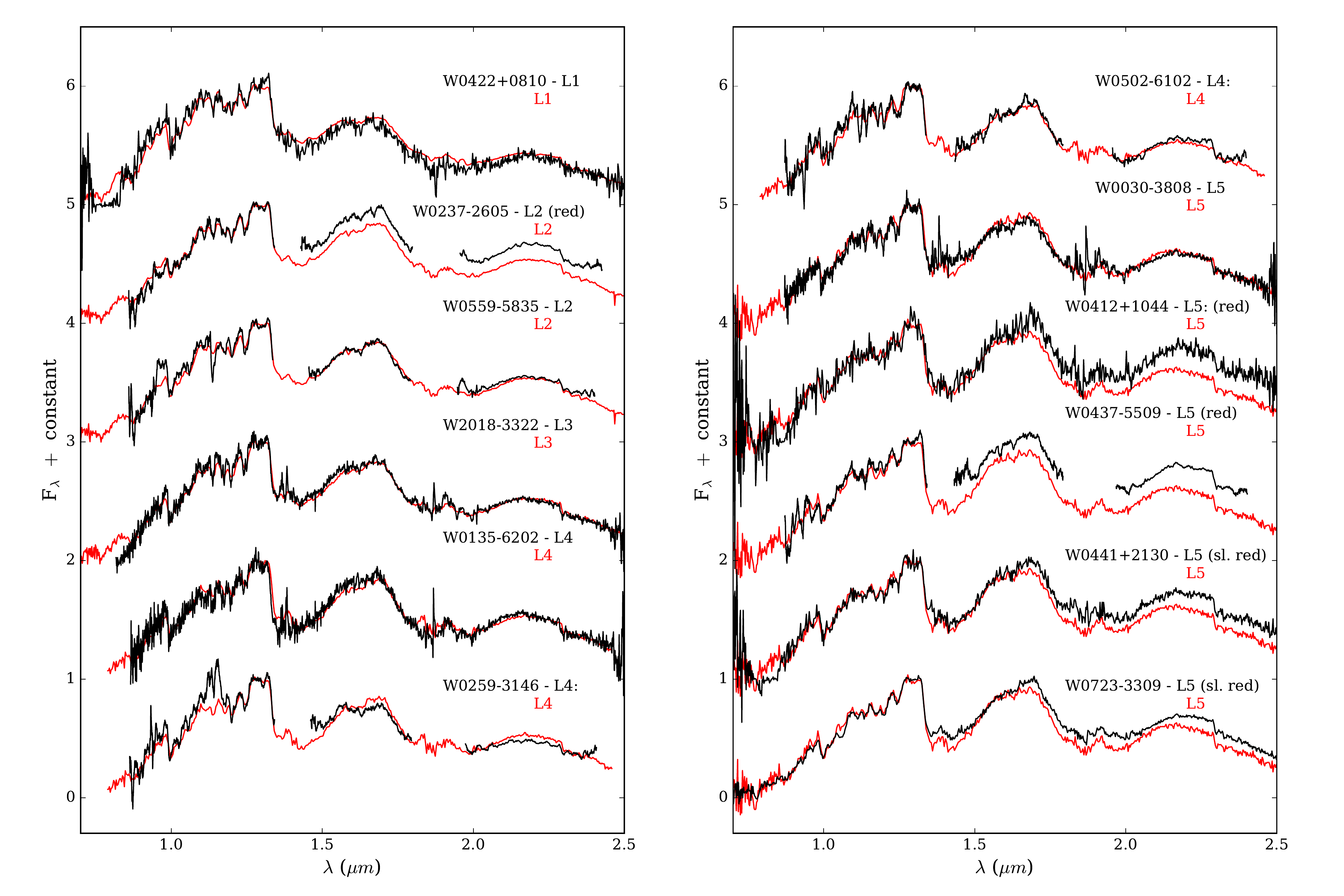}
\caption{Near-infrared spectra of discoveries from this survey.  The object spectra are plotted in black, while the closest matching near-infrared spectral standard is shown in red.  All spectra are normalized between 1.27 and 1.29 $\mu$m.  The standards used for comparison are as follows; L1 -- 2MASSW J2130446$-$084520 \citep{kirk10}, L2 -- Kelu-1 \citep{burg07a}, L3 -- 2MASSW J1506544$+$132106 \citep{burg07b}, L4 -- 2MASS J21580457$-$1550098 \citep{kirk10}, L5 -- SDSS J083506.16$+$195304.4 \cite{chiu06}, L6 -- 2MASSI J1010148$-$040649 \citep{reid06a}, L7 -- 2MASSI J0103320$+$193536 \citep{cruz04}, L8 -- 2MASSW J1632291$+$190441 \citep{burg07b}, L9 -- DENIS-P J0255$-$4700 \citep{burg06}.}  
\end{figure*}

\begin{figure*}
\plotone{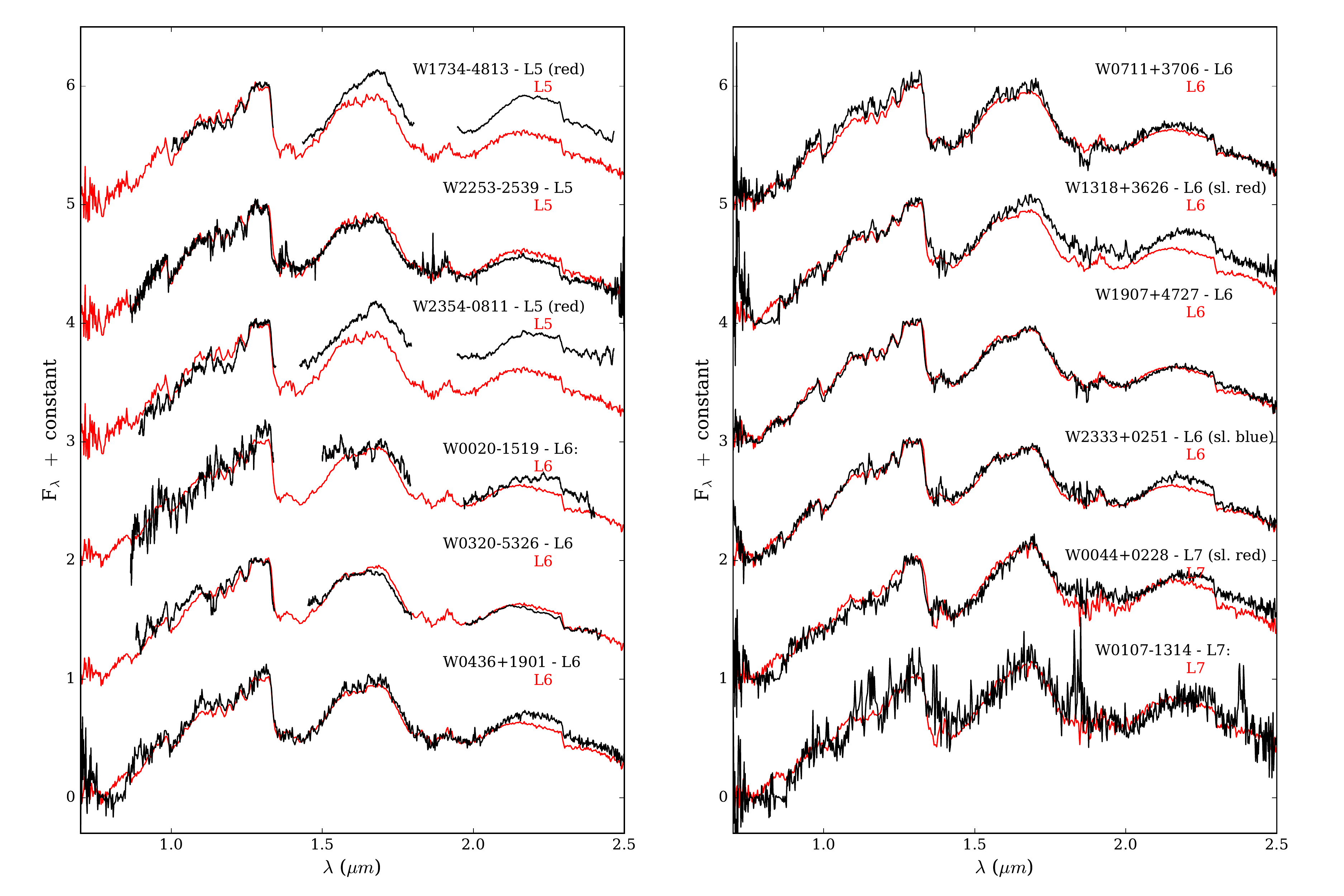}
\caption{Same as Figure 2.}  
\end{figure*}

\begin{figure*}
\plotone{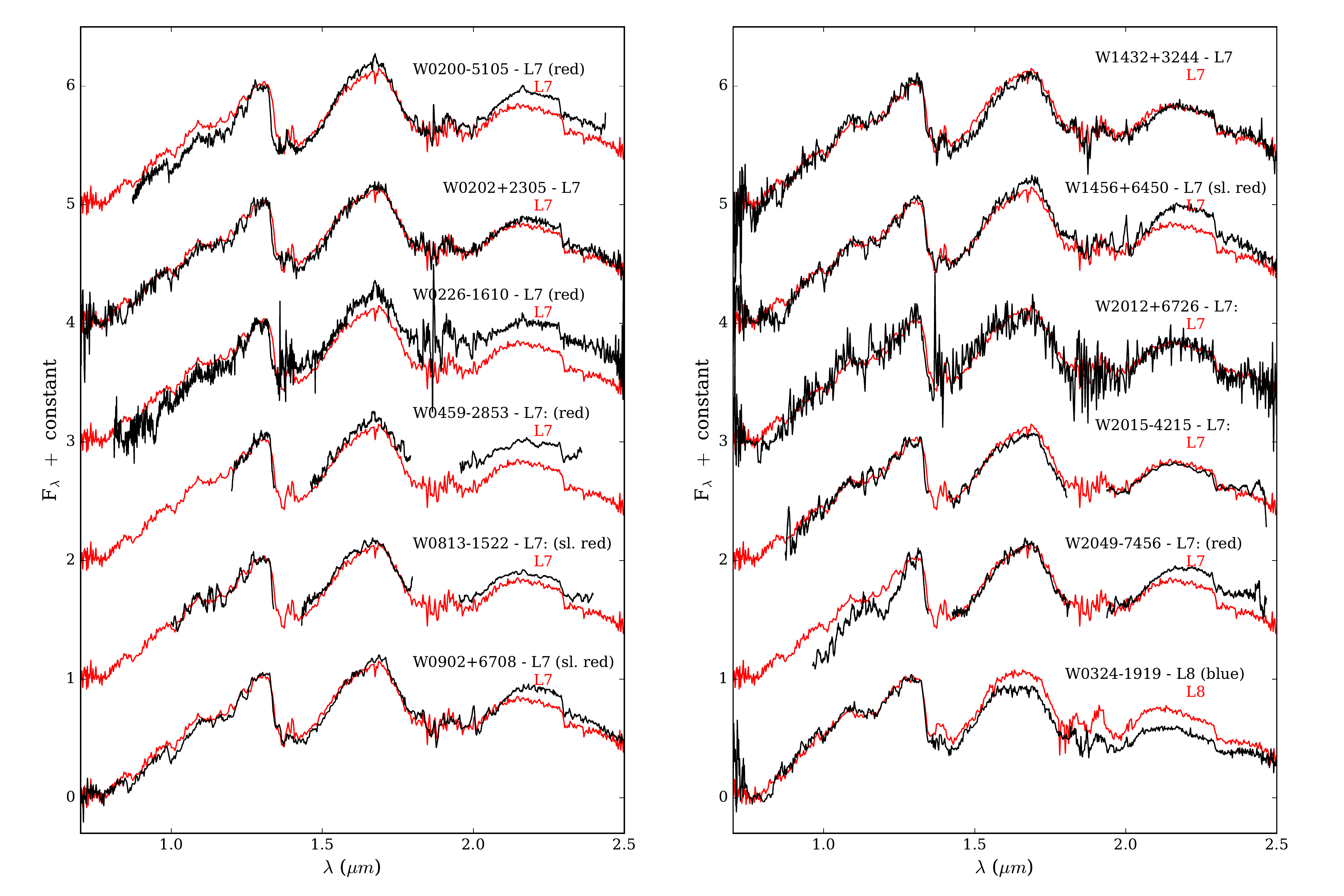}
\caption{Same as Figure 2.}  
\end{figure*}

\begin{figure*}
\plotone{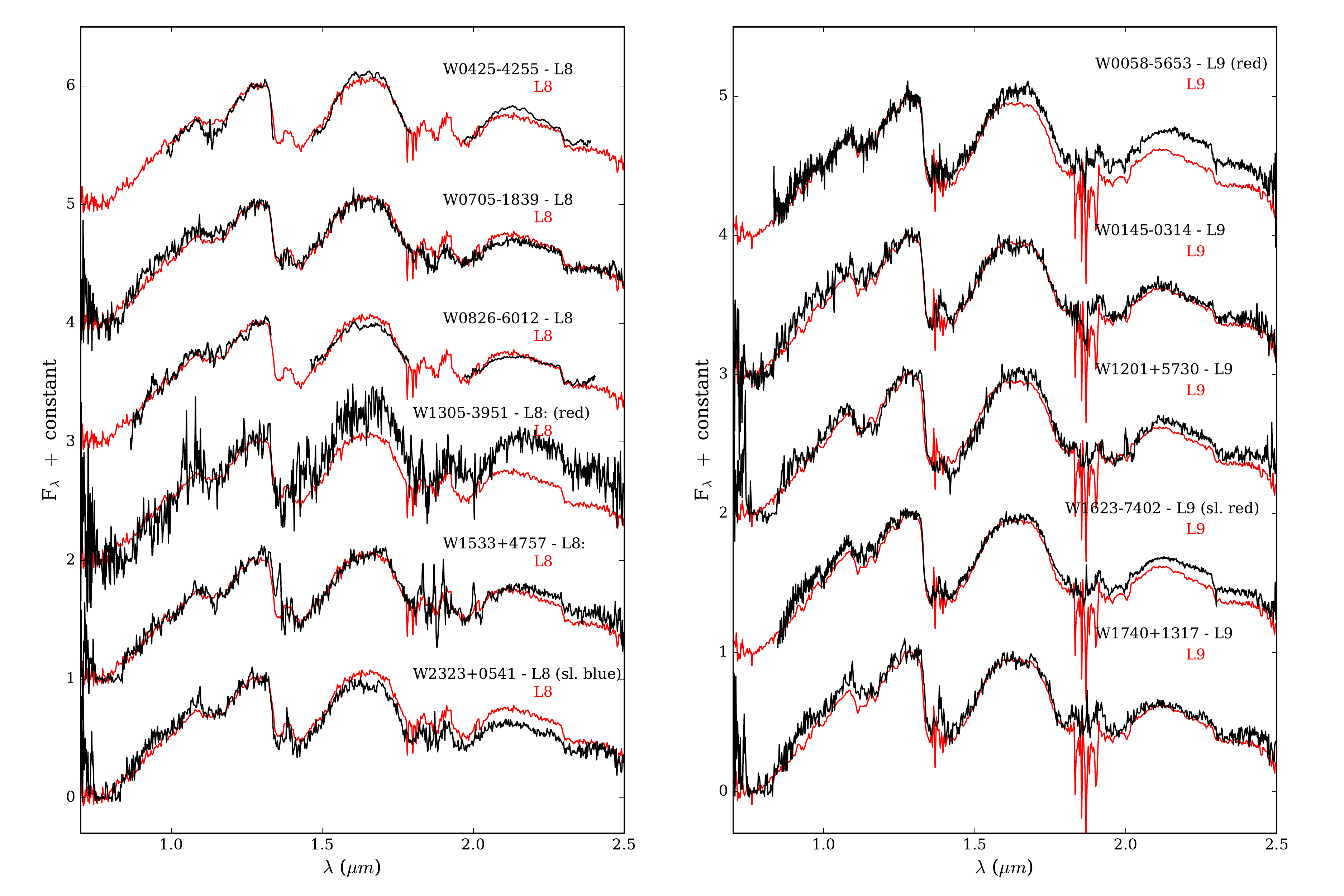}
\caption{Same as Figure 2.}  
\end{figure*}

As discussed in \cite{schneid16b} (see also \citealt{fah13}), the photometric distances of young, late-type L dwarfs closely match their measured parallactic distances using $K$-band magnitudes.  While not all objects in this sample are young, we find that several are likely members of nearby moving groups (see Section 5).  We therefore calculate photometric distances using the 2MASS $K_S$-band magnitude and the absolute magnitude-spectral type relations from \cite{dup12}.  The photometric distance uncertainties include both spectral type and photometric uncertainties.  Two recent, extensive parallax programs focused on young brown dwarfs (\citealt{fah16} and \citealt{liu16}) allow us to investigate if this $K$-band assumption holds true for a much larger sample of low gravity objects than that investigated in \cite{schneid16b}.  If the difference between photometric $K$-band estimates and actual measured parallaxes is small for young brown dwarfs, then there should be little difference between the absolute magnitudes of young L dwarfs and field L dwarfs per spectral type bin when using $K$-band photometry.  Indeed this is the case in both studies (see Figure 17 of \citealt{fah16} and Figure 6 of \citealt{liu16}).  All photometric distance estimates for the new discoveries presented here are given in Table 4.  We assume each object is single for these estimates.  

\begin{deluxetable*}{lcccccccccccc}
\tablecaption{Derived Properties of New Discoveries}
\tablehead{
\colhead{AllWISE Designation} & \colhead{Sp.\ Type} & \colhead{$\mu$$_{\alpha}$} & \colhead{$\mu$$_{\delta}$} & \colhead{$d_{\rm phot}$\tablenotemark{a}}  \\
 & & (mas yr$^{-1}$) & (mas yr$^{-1}$) & (pc)  }
\startdata
J002050.25$-$151913.1 & L6: & 158.5 $\pm$ 15.9 & $-$13.6 $\pm$ 14.9 & 36 $\pm$ 6  \\
J003052.08$-$380829.6 & L5 & 150.1 $\pm$ 26.1 & $-$33.4 $\pm$ 24.4 & 45 $\pm$ 6  \\
J004403.39$+$022810.6 & L7 (sl.\ red) & 104.8 $\pm$ 15.4 & $-$61.9 $\pm$ 14.5 & 31 $\pm$ 3  \\
J005811.69$-$565332.1 & L9 (red) & 197.4 $\pm$ 12.8 & 46.0 $\pm$ 11.9 & 22 $\pm$ 2 \\
J010738.75$-$131413.7 & L7: & 101.3 $\pm$ 11.6 & $-$25.1 $\pm$ 10.8 & 28 $\pm$ 5  \\
J013556.99$-$620245.5 & L4 & 180.6 $\pm$ 29.4 & 91.5 $\pm$ 27.6 & 49 $\pm$ 7 \\
J014535.23$-$031412.9 & L9 & 29.5 $\pm$ 14.9 & $-$97.4 $\pm$ 14.9 & 27 $\pm$ 3  \\
J020047.29$-$510521.4 & L7 (red) & 171.2 $\pm$ 9.9 & $-$75.4 $\pm$ 8.2 & 20 $\pm$ 2  \\
J020229.29$+$230513.9 & L7 & 128.6 $\pm$ 24.7 & $-$15.4 $\pm$ 24.7 & 36 $\pm$ 5  \\
J022609.16$-$161000.4 & L7 (red) & 100.3 $\pm$ 11.5 & $-$108.5 $\pm$ 9.9 & 27 $\pm$ 3 \\
J023749.81$-$260543.8 & L2 (red) & $-$71.0 $\pm$ 13.5 & $-$55.8 $\pm$ 11.1 & 56 $\pm$ 8  \\
J025954.88$-$314655.6 & L4: & $-$227.6 $\pm$ 24.0 & $-$179.7 $\pm$ 24.0 & 57 $\pm$ 13 \\
J032049.31$-$532656.7 & L6 & 80.8 $\pm$ 16.2 & $-$166.8 $\pm$ 14.4 & 35 $\pm$ 4  \\
J032440.23$-$191905.6 & L8 (blue) & $-$118.4 $\pm$ 12.9 & $-$213.3 $\pm$ 12.0 & 25 $\pm$ 3  \\
J041232.77$+$104408.3 & L5: (red) & 114.3 $\pm$ 30.3 & 52.5 $\pm$ 29.4 & 46 $\pm$ 6 \\
J042231.34$+$081012.7 & L1 & $-$115.7 $\pm$ 29.7 & $-$102.0 $\pm$ 28.1 & 80 $\pm$ 12  \\
J042506.66$-$425509.6 & L8 & $-$133.9 $\pm$ 8.6 & $-$100.4 $\pm$ 8.6 & 23 $\pm$ 2   \\
J043642.75$+$190134.8 & L6 & 98.8 $\pm$ 16.5 & $-$29.0 $\pm$ 12.5 & 35 $\pm$ 4 \\
J043718.77$-$550944.0 & L5 (red) & 76.2 $\pm$ 12.9 & 91.4 $\pm$ 12.0 & 35 $\pm$ 7 \\
J044105.56$+$213001.5 & L5 (sl.\ red) & 98.1 $\pm$ 27.3 & $-$39.0 $\pm$ 23.2 & 45 $\pm$ 9  \\
J045900.42$-$285338.3 & L7: (red) & 85.3 $\pm$ 34.4 & 110.3 $\pm$ 30.9 & 38 $\pm$ 8  \\
J050259.73$-$610206.1 & L4: & 35.2 $\pm$ 22.6 & $-$253.3 $\pm$ 20.1 & 47 $\pm$ 11  \\
J055959.30$-$583546.0 & L2 & $-$52.0 $\pm$ 9.9 & $-$99.8 $\pm$ 9.1 & 52 $\pm$ 6  \\
J065935.80$+$771457.8 & \dots & 7.9 $\pm$ 13.2 & $-$146.2 $\pm$ 12.4 & \dots  \\
J070534.00$-$183925.6 & L8 & $-$243.2 $\pm$ 12.4 & 112.7 $\pm$ 11.0 & 26 $\pm$ 2  \\
J071138.88$+$370601.0 & L6 & $-$73.7 $\pm$ 11.9 & $-$287.3 $\pm$ 11.2 & 35 $\pm$ 4  \\
J072352.62$-$330943.5 & L5 (sl.\ red) & $-$37.3 $\pm$ 6.4 & 101.1 $\pm$ 6.4 & 23 $\pm$ 2  \\
J081322.19$-$152203.2 & L7: (sl.\ red) & $-$164.5 $\pm$ 14.7 & 98.8 $\pm$ 13.8 & 31 $\pm$ 6  \\
J082624.09$-$601202.8 & L8 & $-$156.9 $\pm$ 16.6 & $-$234.2 $\pm$ 14.0 & 28 $\pm$ 3  \\
J090258.99$+$670833.1 & L7 (sl.\ red) & $-$112.7 $\pm$ 10.6 & $-$213.2 $\pm$ 9.8 & 23 $\pm$ 2  \\
J093858.10$+$761211.5 & \dots & $-$12.6 $\pm$ 15.4 & 104.2 $\pm$ 14.6 & \dots  \\
J120104.57$+$573004.2 & L9 & 98.6 $\pm$ 28.9 & 13.0 $\pm$ 25.6 & 30 $\pm$ 3  \\
J130523.06$-$395104.9 & L8: (red) & $-$241.7 $\pm$ 15.8 & $-$49.4 $\pm$ 15.8 & 30 $\pm$ 5  \\
J131845.58$+$362614.0 & L6 (sl.\ red) & $-$88.2 $\pm$ 23.1 & 23.4 $\pm$ 20.0 & 40 $\pm$ 5  \\
J143211.17$+$324433.8 & L7 & $-$116.5 $\pm$ 25.5 & 1.8 $\pm$ 24.6 & 35 $\pm$ 4  \\
J145642.68$+$645009.7 & L7 (sl.\ red) & $-$198.8 $\pm$ 13.3 & 54.8 $\pm$ 12.4 & 30 $\pm$ 3  \\
J153358.52$+$475706.9 & L8: & $-$135.8 $\pm$ 20.9 & 35.0 $\pm$ 20.9 & 29 $\pm$ 4  \\
J162341.27$-$740230.4 & L9 (sl.\ red) & $-$122.6 $\pm$ 16.8 & $-$378.7 $\pm$ 14.0 & 26 $\pm$ 3  \\
J173453.90$-$481357.9 & L5 (red) & $-$100.6 $\pm$ 9.3 & $-$228.5 $\pm$ 9.2 & 25 $\pm$ 2  \\
J174057.82$+$131709.4 & L9 & $-$17.5 $\pm$ 21.5 & $-$220.4 $\pm$ 21.6 & 30 $\pm$ 4  \\
J190722.56$+$472745.3 & L6 & 40.1 $\pm$ 8.4 & $-$89.6 $\pm$ 8.3 & 27 $\pm$ 3  \\
J201204.11$+$672608.0 & L7: & 186.1 $\pm$ 25.2 & 216.9 $\pm$ 24.3 & 35 $\pm$ 6  \\
J201530.67$-$421542.5 & L7: & 76.5 $\pm$ 35.1 & $-$55.1 $\pm$ 33.3 & 39 $\pm$ 7  \\
J201826.00$-$332207.3 & L3 & 26.6 $\pm$ 22.1 & $-$186.8 $\pm$ 21.3 & 61 $\pm$ 9  \\
J204902.80$-$745613.5 & L7: (red) & 45.9 $\pm$ 29.5 & $-$120.8 $\pm$ 27.5 & 37 $\pm$ 5  \\
J225333.00$-$253948.0 & L5 & 145.9 $\pm$ 23.1 & $-$40.7 $\pm$ 20.6 & 43 $\pm$ 6  \\
J232307.08$+$054113.0 & L8 (sl.\ blue) & $-$179.8 $\pm$ 32.7 & $-$112.8 $\pm$ 30.7 & 38 $\pm$ 5  \\
J232453.73$+$503525.4 & \dots & 182.0 $\pm$ 11.3 & 58.2 $\pm$ 11.2 & \dots  \\
J233333.46$+$025128.4 & L6 (sl.\ blue) & 272.7 $\pm$ 14.4 & 169.9 $\pm$ 10.1 & 32 $\pm$ 3  \\
J235422.31$-$081129.7 & L5 (red) & 130.1 $\pm$ 14.2 & $-$70.1 $\pm$ 12.6 & 38 $\pm$ 5 \\
\enddata
\tablenotetext{a}{Photometric distance using the 2MASS $K_S$ magnitude and the absolute magnitude-spectral type relations from \cite{dup12}.}
\end{deluxetable*}

\subsection{Sample Comparison}

We first inspect all objects recovered with this survey as a whole.  The left panel of Figure 6 shows a color-color diagram comparing the positions of four distinct samples recovered in this survey; known young L dwarfs (``Y'' or ``Y?'' in Table 1), known L dwarfs without any mention of youth in the literature, new discoveries from this survey with ``red'' or ``sl.\ red'' spectral types, and the remainder of the discoveries from this survey.  The known young L dwarfs and new ``red'' discoveries generally occupy the same region of color space.  While the new discoveries not labeled as ``red'' extend over much of the same color space as known young Ls and the new ``red'' discoveries, there is a significantly larger population of these objects at the bluest corner of this color space.  This is not unexpected as the vast majority of these objects are expected to have ages consistent with the field population.

We also compare the same sets of objects on a reduced proper motion diagram to investigate whether these populations show kinematically distinct characteristics.  The reduced proper motion is defined as H$_m$ = $m$ $+$ 5log($\mu$) $+$ 5, where $m$ is a particular photometric band and $\mu$ is the object's total proper motion.  We chose to use the 2MASS $K_{\rm S}$ magnitude for this diagram as some objects have upper limits in the 2MASS $J$ band.  Again, the new ``red'' discoveries and the known young L dwarfs generally occupy the same regions of this diagram.  While there is some overlap, these two groups look to be kinematically distinct from both the other known L dwarfs and discoveries not found to be ``red''.  In general, the new ``red'' discoveries and known young L dwarfs tend to be redder and have smaller reduced proper motion magnitudes.  While seventeen of the 47 known young or new ``red'' objects have reduced proper motion magnitudes greater than 16 ($\sim$36\%), a significantly larger fraction of the remaining objects have reduced proper motion magnitudes greater than 16 (30/50, or 60\%).  

\begin{figure*}
\plotone{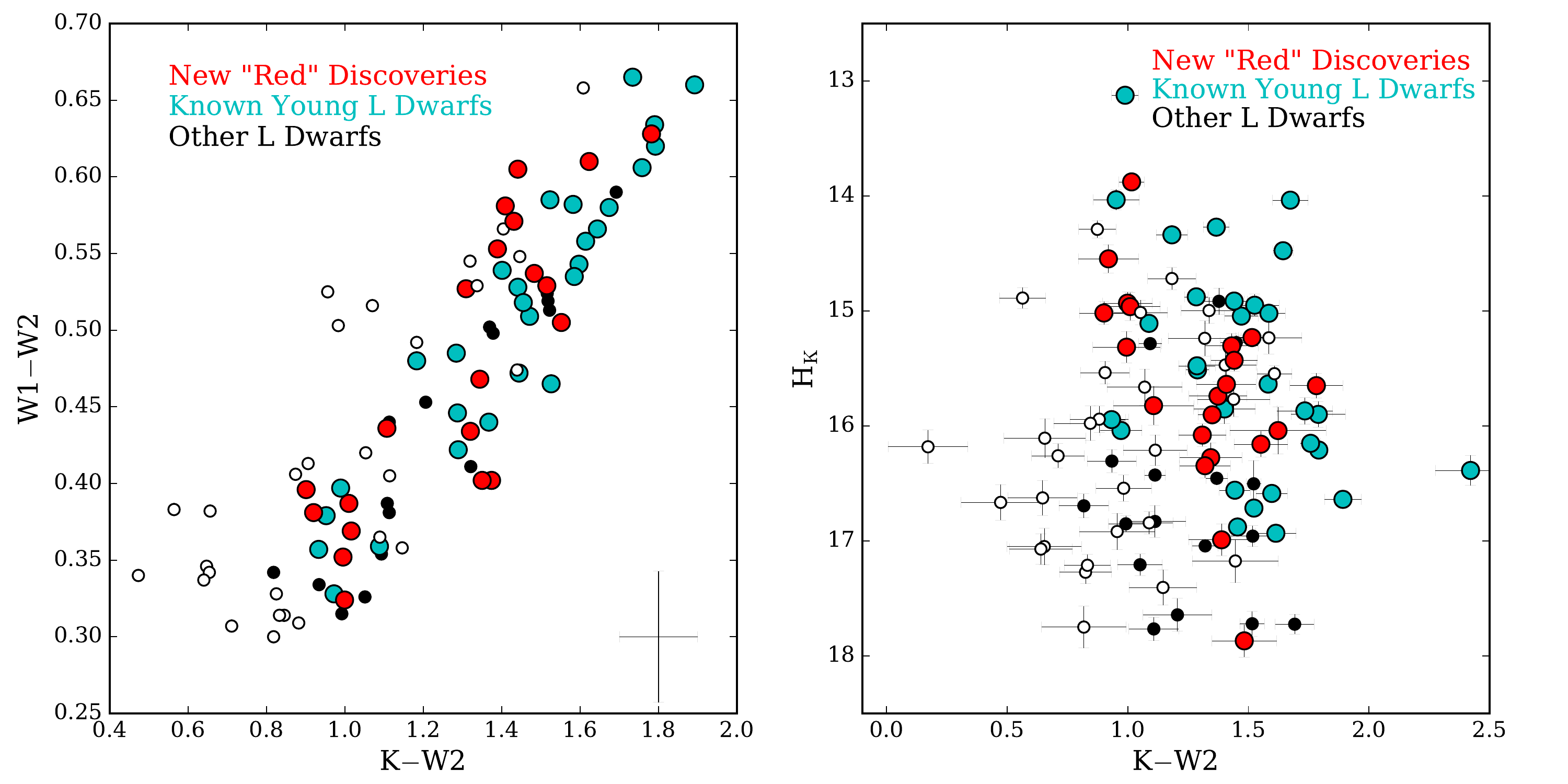}
\caption{{\it Left:} Color-color diagram comparing the discoveries from this survey to previously known young brown dwarfs.  The typical uncertainty for each plotted symbol is shown in the bottom right corner.  {\it Right:}  Reduced proper motion diagram comparing the discoveries from this survey to previously known young brown dwarfs. The discoveries with ``red'' or ``sl.\ red'' spectral types are shown at red symbols, while known young L dwarfs are shown in cyan.  All other plotted objects have no signs of youth in their spectra, both new discoveries (open symbols) and previously known brown dwarfs (filled black symbols).  WISEA J114724.10$-$204021.3, which was discovered as part of this survey but published separately \citep{schneid16b} is plotted as a red symbol. }  
\end{figure*}

\subsection{Previously Proposed Brown Dwarf Candidates}

There are several objects which have been identified previously as brown dwarf candidates in the literature that we provide the first spectra of here.

\subsubsection{WISEA J004403.39$+$022810.6, WISEA J232307.08$+$054113.0, and WISEA J233333.46$+$025128.4}

All three of these objects were identified as a brown dwarf candidates in \cite{skrz16}.   \cite{skrz16} classified these objects using photometry alone and found L7p, L8.5, and L4.5 for WISEA J004403.39$+$022810.6, WISEA J232307.08$+$054113.0, and WISEA J233333.46$+$025128.4, respectively. We find similar spectral types from our spectra in all three instances; L7 (sl. red) for WISEA J004403.39$+$022810.6, L8 (sl.\ blue) for WISEA J232307.08$+$054113.0, and L6 for WISEA J233333.46$+$025128.4.  See additional discussion of the potential youth and moving group membership of WISEA J004403.39$+$022810.6 in Section 5.1.1. 

\subsubsection{WISEA J005811.69$-$565332.1 and WISEA J020047.29$-$510521.4}

These objects were identified as a potential moving group members in \cite{gagne15a}, where WISEA J005811.69$-$565332.1 was found to be a modest probability Argus member and WISEA J020047.29$-$510521.4 was identified as a high probability ($\sim$99\%) member of ABDor.  We present the first spectroscopic confirmation of these objects as L dwarfs in Figures 4 and 5 and determine a spectral types of L7 (sl.\ red) and L9 (red) for WISEA J005811.69$-$565332.1 and WISEA J020047.29$-$510521.4, respectively.  Additional discussion of the potential youth and moving group membership of these objects is provided in Sections 5.1.2 and 5.1.3. 

\subsubsection{WISEA J162341.27$-$740230.4 and WISEA J201204.11$+$672608.0}

These objects were previously identified as high proper motion objects in \cite{luh14}.  We present the first spectrum of these objects in Figures 4 and 5 and determine spectral types of L9 (sl.\ red) and L7: for WISEA J162341.27$-$740230.4 and WISEA J201204.11$+$672608.0, respectively. See additional discussion of the potential youth and moving group membership of WISEA J162341.27$-$740230.4 in Section 5.1.17.

\section{Potentially Young Objects}

A significant portion of our new discoveries show redder spectral slopes compared to their corresponding near-infrared spectral standards, and are labeled as ``red'' or ``sl. red'' in Table 4.  For these potentially young objects (20 total), we investigate whether or not they belong to any young, nearby, moving groups with three different available algorithms; BANYAN II (\citealt{malo13}, \citealt{gagne14}), LACEwING \citep{ried17}, and the Convergent Point (``CP'') method from \cite{rod13}.  For BANYAN II and LACEwING, we use the ``young'' option for all objects.  

The BANYAN II moving group membership evaluation tool uses a naive Bayesian classifier analysis to assess membership probabilities for several nearby, young moving groups, including the AB Doradus moving group (AB Dor), Argus, the $\beta$ Pictoris moving group ($\beta$ Pic), Carina, Columba, Tucana-Horologium (Tuc-Hor), and the TW Hydra association (TWA).  For each suspected young object in our sample, we use its proper motion and AllWISE position to determine potential membership probabilities.  In Table 5, we report all membership probabilities greater than 10\% found with BANYAN II.  We also report the predicted distances given by BANYAN II assuming an object is a moving group member.  Note that \cite{liu16} find that the kinematic distances from BANYAN II agree to within 2.3$\sigma$ with measured parallax distances for 90\% of the objects they compared.  In almost all cases where photometric and predicted distances disagree, the predicted distance is significantly closer than the calculated photometric distance.  In such cases, the question of unresolved binarity would only increase photometric distance estimates, making such discrepancies worse.  

The convergent point analysis tool of \cite{rod13} uses positions and proper motions to determine motions relative to the convergent points of 6 different groups; AB Dor, $\beta$ Pic, Carina-Near, Columba, Tuc-Hor, and TWA.  We use AllWISE positions and the proper motions of each suspected young object as inputs into the convergent point tool and list the results in Table 5.  We include matches with probabilities higher than 50\%.

In addition to the groups listed above for BANYAN II and the convergent point tool, the LACEwING kinematic analysis tool also evaluates membership for several additional groups, including $\eta$ Cha, $\epsilon$ Cha, 32 Ori, Octans, Coma Ber, Ursa Major, $\chi$$^{01}$ For, and the Hyades.  Again, we use AllWISE positions and the proper motions listed in Table 4 to evaluate each potentially young object with LACEwING and provide the results in Table 5.  All matches with probabilities $>$20\% are provided.  We note, however, that the LACEwING tool seems to be much more efficient for objects with complete kinematic information (position, proper motion, radial velocity, and parallax), while our candidates only have positions and proper motions.  For example, for our newly discovered L5 (red) dwarf WISEA J235422.31$-$081129.7, BANYAN II finds a 94\% probability of belonging to $\beta$ Pic and the convergent point tool finds a probability of 100\%, whereas LACEwING finds a 0\% probability of belonging to $\beta$ Pic.  However, if we input the predicted distance and radial velocity for WISEA J235422.31$-$081129.7 from BANYAN II into LACEwING, the LACEwING code finds a 60\% probability of belonging to $\beta$ Pic.  Thus, for moving group membership evaluation, we rely primarily on results from BANYAN II and the convergent point tool and include the results from LACEwING for completeness.  We list, at most, two matching associations per object per moving group evaluation tool.

\subsection{Notes on Individual Objects}

\subsubsection{WISEA J004403.39$+$022810.6}

This object has a high probability of belonging to $\beta$ Pic according to both BANYAN II ($\sim$78\%) and the convergent point tool ($\sim$97\%).  Its photometric distance (31 $\pm$ 3 pc) matches very well with the distance predicted by BANYAN II (33 $\pm$ 4 pc) and the convergent point tool (36 pc).  Note that while the convergent point tool also finds a high probability of belonging to Columba ($\sim$99\%), its predicted distance if a Columba member (46 pc) is significantly different from its photometric distance.  

Its spectrum shows a peaked $H$-band shape, a common feature among very young brown dwarfs, and a redder spectral shape compared to the L7 near-infrared spectral standard.  There are two spectral indices that have been shown to be effective for distinguishing young, low-gravity brown dwarfs from the field population for spectral types later than L4; the $H$-cont index \citep{allers13} and the H2($K$) index (\citealt{canty13}, \citealt{schneid14}).  Both indices probe areas where the effects of collisionally induced absorption of H$_2$ is greatly reduced in low-gravity objects, affecting their overall spectral shape.  However, we note that \cite{allers13} caution that the use of the $H$-cont index is not the most reliable gravity indicator, and should be used in combination with other indices when possible.  For WISEA J004403.39$+$022810.6, we find an $H$-cont value of 0.968 and H2($K$) of 1.009, both of which align with other low-gravity objects with similar spectral types.  We thus conclude that this object is a high probability member of $\beta$ Pic.  Radial velocity and parallax measurements will be needed to confirm membership.  If a true $\beta$ Pic member, WISEA J004403.39$+$022810.6 would be one of the latest spectral type members known, and would hence have a very low mass.  Using its photometric distance estimate, an age of 24 $\pm$ 3 Myr \citep{bell15}, the spectral type-$K_{\rm S}$ bolometric correction relation for young objects from \cite{fil15}, and the evolutionary models of \cite{sau08}, we find a mass range of 7$-$11 $M_{\rm Jup}$, which would place WISEA J004403.39$+$022810.6 in the planetary mass regime.  

While \cite{allers13} provide a suggested list of low and intermediate gravity standards, they did not find suitable standards for spectral type L7.  We thus compare the near-infrared spectrum of WISEA J004403.39$+$022810.6 to the low and intermediate surface gravity L6 standards in Figure 7.  

\subsubsection{WISEA J005811.69$-$565332.1}

This object is a modest probability member of both $\beta$ Pic ($\sim$27\%) and Argus ($\sim$31\%) according to our input into BANYAN II, and has a modest probability of belonging to TucHor ($\sim$48\%) and ABDor ($\sim$40\%) according to LACEwING.  Its photometric distance (22 $\pm$ 3 pc) is similar to and in between the predicted distances for both BANYAN II matched associations (19 $\pm$ 2 pc for $\beta$ Pic and 25 $\pm$ 3 pc for Argus).  The predicted distances for the LACEwING matches are slightly more discrepant (29 $\pm$ 3 and 29 $\pm$ 2 for TucHor and ABDor, respectively).  The effects of low surface gravity on the spectral features of L9 type dwarfs has yet to be thoroughly explored, thus we cannot comment further on the youth of this object from the available spectrum.  We note that a radial velocity measurement could help clear up the ambiguous moving group membership of this object, as the predicted radial velocities are 9.6, 2.5, 7.2, and 21.9 km s$^{-1}$ for $\beta$ Pic, Argus, TucHor, and ABDor, respectively.     

\subsubsection{WISEA J020047.29$-$510521.4}
 
This object is a high probability member of ABDor according our input into BANYAN II ($\sim$98\%), and a moderate probability ABDor member according to the convergent point tool ($\sim$54\%) and LACEwING ($\sim$53).  Its photometric distance estimate (20 $\pm$ 2 pc) matches very well with its predicted distance if it is an ABDor member for all membership tools;  22 $\pm$ 2 pc for BANYAN II, 23 $\pm$ 2 for LACEwING, and 23 pc for the convergent point tool.  Furthermore, its spectrum shows a redder than normal near-infrared shape compared to spectral standards as it's one of the reddest objects in our sample ($J-K_S$ = 2.54 mag).  We find $H$-cont = 0.904 and H2($K$) = 1.067, values near the boundary between field and low-gravity objects.  This is consistent with the age of ABDor ($\sim$149 Myr; \citealt{bell15}).  We thus conclude that this object is a high likelihood member of ABDor.  Radial velocity and parallax measurements will be needed to confirm membership.  Using the spectral type-$K_{\rm S}$ bolometric correction relation from \cite{fil15}, evolutionary models from \cite{sau08}, an age of 149$^{+51}_{-19}$ \citep{bell15}, and WISEA J020047.29$-$510521.4's distance estimate, we find a mass range of 16-28 $M_{\rm Jup}$ for this object.  A comparison of the near-infrared spectrum of WISEA J020047.29$-$510521.4 to the low and intermediate surface gravity L6 standards in shown in Figure 7.   

\subsubsection{WISEA J022609.16$-$161000.4}

According to BANYAN II, this object has a high probability of belonging to ABDor ($\sim$85) as well as a small probability of belonging to $\beta$ Pic ($\sim$12\%).  The convergent point tool returns a very high probability of belonging to ABDor ($\sim$99\%), while LACEwING returns modest probabilities for both ABDor ($\sim$36\%) and TucHor ($\sim$36\%).  The photometric distance estimate for this object (27 $\pm$ 3 pc) is consistent within 3$\sigma$ for all three ABDor  distance estimates (36 $\pm$ 3 pc, 37 $\pm$ 1 pc, and 36 pc).  Its near-infrared spectrum shows a peaky $H$-band and a red slope compared to the L7 standard.  We measure $H$-cont = 0.989 and H2($K$) = 1.036, values consistent with other low-gravity objects.  This object is also one of the reddest  objects in our sample ($J-K_S$ = 2.74 mag).  We consider this object a high likelihood member of ABDor.  If a true member, we find a mass range of 16-28 $M_{\rm Jup}$ using the spectral type-$K_{\rm S}$ bolometric correction relation from \cite{fil15}, evolutionary models from \cite{sau08}, an age of 149$^{+51}_{-19}$ \citep{bell15}, and WISEA J022609.16$-$161000.4's distance estimate.   We compare the near-infrared spectrum of WISEA J022609.16$-$161000.4 to  low and intermediate surface gravity L6 standards in Figure 7.

\subsubsection{WISEA J023749.81$-$260543.8}

This object's spectrum is significantly redder than the L2 near-infrared spectral standard.  This object is not a likely member of any nearby group evaluated by BANYAN II, LACEwING, or the convergent point tool.  Because it has a spectral type of $\sim$L2, we measure all low-gravity indices found in \cite{allers13} and find values consistent with a field age population.  The origin of the red near-infrared colors of this object is unknown. 

\subsubsection{WISEA J041232.77$+$104408.3}

This object is red compared to the L5 near-infrared spectral standard, but does not belong to any nearby group according to BANYAN II  and the convergent point tool.  LACEwING, however, finds a high probability of belonging to the Hyades cluster ($\sim$87\%).  This object's photometric distance estimate (46 $\pm$ 6 pc) is fully consistent with Hyades membership, though its $\mu_{\delta}$ value is somewhat discrepant with other Hyades members.  Of the 724 Hyades members in \cite{roe11}, only 13 have positive $\mu_{\delta}$ values as large as WISEA J041232.77$+$104408.3, all of which have negative declinations.  While this does not rule out Hyades membership for this object, we consider it an unlikely member.   

\subsubsection{WISEA J043642.75$+$190134.8 and WISEA J044105.56$+$213001.5}

These objects have sky positions coincident with the Hyades cluster.  The LACEwING tool finds Hyades membership probabilities of $\sim$87\% and 100\% for WISEA J043642.75$+$190134.8 and WISEA J044105.56$+$213001.5, respectively.  The photometric distance estimate of WISEA J044105.56$+$213001.5 (45 $\pm$ 9 pc) is well within the Hyades tidal radius of 9 pc \citep{roe11} from a nominal Hyades distance of $\sim$47 pc \citep{van09}, and is fully consistent with the predicted LACEwING distance of 46 $\pm$ 6 pc.  The photometric distance estimate of WISEA J043642.75$+$190134.8 (35 $\pm$4 pc) is 1$\sigma$ from the edge of the tidal radius, and is within 2$\sigma$ of the LACEwING predicted distance of 47 $\pm$ 5 pc. The average $\mu_{\alpha}$ and $\mu_{\delta}$ values for Hyades members from \cite{roe11} is 104.9 mas yr$^{-1}$ and -27.3 mas yr$^{-1}$, respectively. We find $\mu_{\alpha}$ = 98.8 $\pm$ 16.5 mas yr$^{-1}$ and $\mu_{\delta}$ = -29.0 $\pm$ 12.5 mas yr$^{-1}$ for WISEA J043642.75$+$190134.8 and $\mu_{\alpha}$ = 98.1 $\pm$ 27.3 mas yr$^{-1}$ and $\mu_{\delta}$ = -39.0 $\pm$ 23.2 mas yr$^{-1}$ for WISEA J044105.56$+$213001.5, both consistent with other Hyades members.  At L5 and L6, these would be the latest spectral type members of the Hyades with the exceptions of CFHT-Hy-20 \citep{bou08}, recently confirmed as a T2 spectral type member in \cite{liu16} and 2MASS J04183483$+$2131275, a recently confirmed L5 Hyades member \citep{perez17}.   

\subsubsection{WISEA J043718.77$-$550944.0}

This object's spectrum is very red compared to the L5 spectral standard, and according to BANYAN II and the convergent point tool, is a high probability member of $\beta$ Pic ($\sim$94\% and $\sim$93\%, respectively). However, the predicted distances (17 $\pm$ 3 pc and 19 pc) do not agree with the estimated photometric distance (35 $\pm$ 7 pc).  We measure an $H$-cont value of 0.975, consistent with having a low-gravity, but measure H2($K$) = 1.064, which is indicative of a field age gravity.  This object is worthy of additional observations to untangle its potential youth and moving group membership.  

\subsubsection{WISEA J045900.42$-$285338.3}

According to BANYAN II, this object has a modest probability of belonging to both $\beta$ Pic ($\sim$29\%) and Argus ($\sim$55\%).  However, the predicted distances in both cases do not agree with the estimated photometric distance. The convergent point tool finds a large probability of belonging to Carina-Near ($\sim$97\%) and a distance within 2$\sigma$ of its photometric distance estimate.  However, we note that the sky position of WISEA J045900.42$-$285338.3 is not near to any of the proposed Carina-Near members in \cite{zuck06}.  We measure $H$-cont = 0.943 and H2($K$) = 1.012, values consistent with having a low surface gravity.  This object may belong to an as yet unknown group.  

\subsubsection{WISEA J072352.62$-$330943.5}

This object has a reasonable probability of belonging to Argus according to BANYAN II ($\sim$53\%) and LACEwING ($\sim$25\%) and its spectrum is redder than the near-infrared L5 standard.  Note that the convergent point tool does not consider the Argus moving group.  Its distance estimate from photometry (23 $\pm$ 2 pc) is in good agreement with the BANYAN predicted distance if an Argus member (26 $\pm$ 4 pc).  We find $H$-cont = 0.923 and H2($K$) = 1.031, values consistent with an intermediate to low surface gravity. We suggest this object is a medium to high probability member of Argus.  If an Argus member, with an age of $\sim$40 Myr \citep{tor08}, then it would have a mass of 11-12 $M_{\rm Jup}$, using its distance estimate, the spectral type-$K_{\rm S}$ bolometric correction relation from \cite{fil15}, and evolutionary models from \cite{sau08}.  

\subsubsection{WISEA J081322.19$-$152203.2}

This object's spectrum is slightly red compared to the L7 near-infrared standard.   Its a high probability member of Argus ($\sim$87\%) according to BANYAN II and Carina-Near ($\sim$97\%) according to the convergent point tool, but its photometric distance estimate (31 $\pm$ 6 pc) is significantly different than the BANYAN II predicted distance (15 $\pm$ 2 pc) and the convergent point tool distance (17 pc).  We find $H$-cont = 0.941, consistent with a low surface gravity, but find H2($K$) = 1.054, which coincides with field age objects.  The age of this object remains ambiguous.

\subsubsection{WISEA J090258.99$+$670833.1}

This object's spectrum is slightly red compared to the L7 near-infrared standard.   It is a fairly high probability member of ABDor ($\sim$56\%) according to BANYAN II and its predicted distance (23 $\pm$ 2 pc) is a perfect match to its photometric distance (23 $\pm$ 2 pc).  However, the convergent point tool find a high probability of belonging to Columba ($\sim$83\%) with a similar distance.  Note that both LACEwING and the convergent point tool find probabilities of belonging to ABDor just below our threshold ($\sim$14\% and $\sim$46\%, respectively) and predict a distance of $\sim$24 pc.  This object is one of the reddest objects in our sample ($J-K_S$ = 2.73 mag).  We measure $H$-cont = 0.918 and H2($K$) = 1.014, values completely consistent with having a low surface gravity.  We conclude this object is a medium to high probability ABDor member.  Using the spectral type-$K_{\rm S}$ bolometric correction relation from \cite{fil15}, evolutionary models from \cite{sau08}, an age of 149$^{+51}_{-19}$ \citep{bell15}, and a distance estimate of 23.4 pc, we find a mass range of 16-28 $M_{\rm Jup}$ if a true ABDor member.  A comparison of the near-infrared spectrum of WISEA J090258.99$+$670833.1 to the low and intermediate surface gravity L6 standards in shown in Figure 7.      

\subsubsection{WISEA J130523.06$-$395104.9}

This object's spectrum is red compared to the near-infrared L8 standard, but is rather noisy, so the spectral type is uncertain.  It has a large probability of belonging to Argus ($\sim$95\%) according to BANYAN II, but its predicted distance (21 $\pm$ 2 pc) differs from its photometric distance estimate (30 $\pm$ 5 pc).  The convergent point tool returns possible membership in TucHor ($\sim$80\%) and Carina-Near ($\sim$59\%), though the distance estimate if a TucHor member (18 pc) and the sky position compared to Carina-Near members from \cite{zuck06} makes membership in either unlikely.  We retain this object as a possible Argus member.  A higher S/N spectrum would help to confirm this object's youth. 

\subsubsection{WISEA J131845.58$+$362614.0}

This object's spectrum is red compared to the L6 spectral standard, but does not seem to belong to any nearby group according to BANYAN II.  The convergent point tool finds a high probability of belonging to Carina-Near ($\sim$90\%), though its distance estimate (72 pc) compared to its photometric distance (40 $\pm$ 5 pc) makes membership unlikely.   We find $H$-cont = 0.930 and H2($K$) = 1.006, values consistent with young, low-gravity objects.  This object may belong to an as yet unknown young, nearby group.

\subsubsection{WISEA J145642.68$+$645009.7}

This object's spectrum is slightly red compared to the L7 near-infrared spectral standard.  It has a good probability of belonging to ABDor according to BANYAN II ($\sim$52\%) and the convergent point tool ($\sim$71\%).  This object is also one of the reddest object in our sample ($J-K_S$ = 2.79 mag).  The photometric distance of this object (30 $\pm$ 3 pc) and predicted distances (16 $\pm$ 2 pc and 17 pc) do not agree, however.  We measure $H$-cont = 0.910 and H2($K$) = 1.044, consistent with an intermediate surface gravity.  The youth of this object remains ambiguous.

\subsubsection{WISEA J162341.27$-$740230.4}

This object has a moderate probability of belong to either $\beta$ Pic or ABDor according to BANYAN II (58\% and 36\%, respectively), Argus and ABDor according to LACEwING ($\sim$32\% and $\sim$38\%, respectively), and ABDor ($\sim$70\%) according to the convergent point tool. However, its photometric distance estimate (26 $\pm$ 3 pc) is much further than the predicted distances for all of these groups, and thus unlikely to be a member of any of them. 

\subsubsection{WISEA J173453.90$-$481357.9}

This object's spectrum is very red compared to the L5 spectral standard.  BANYAN II gives that this object may belong to Argus ($\sim$45\%), but its photometric distance (25 $\pm$ 2 pc) does not match well with its predicted distance (13 $\pm$ 2 pc).  LACEwING also suggests possible Argus membership ($\sim$36\%) as well as ABDor ($\sim$24\%), while the convergent point tool suggests membership in Carina-Near ($\sim$63\%).  The predicted ABDor distance from LACEwING (28 $\pm$ 2 pc) is consistent with this object's photometric distance.  However, because both BANYAN II and the convergent point tool give ABDor probabilities of 0\%, we consider this object a low-probability ABDor member. Its measured values of $H$-cont = 0.939 and H2($K$) = 1.032 suggest an intermediate surface gravity.  

\subsubsection{WISEA J204902.80$-$745613.5}

This object is redder than the L7 spectral standard and has a moderate probability of belonging to TucHor according to BANYAN II ($\sim$42\%), LACEwING ($\sim$31\%), and the convergent point tool ($\sim$81\%).  While its photometric distance estimate of 37 $\pm$ 5 pc disagrees with the predicted distance from BANYAN II for TucHor (49 $\pm$ 5 pc), it agrees almost perfectly with the predicted distances from LACEwING (37 $\pm$ 7 pc) and the convergent point tool (37 pc).  We thus conclude that this is a moderate-probability TucHor member worthy of additional follow up observations.  If confirmed as at TucHor member, this would be the latest spectral type member known.  We measure $H$-cont = 0.938 and H2($K$) = 1.010, consistent with a low surface gravity.  We compare the near-infrared spectrum of WISEA J204902.80$-$745613.5 to  low and intermediate surface gravity L6 standards in Figure 7.  

\subsubsection{WISEA J235422.31$-$081129.7}

This object is very red compared to the L5 spectral standard.  It is a high probability member of $\beta$ Pic according to both BANYAN II ($\sim$94\%) and the convergent point too (100\%).  Its photometric distance (38 $\pm$ 5) is within 2$\sigma$ of its predicted distance in both instances (29 $\pm$ 3 pc and 30 pc).  We measure $H$-cont = 0.991 and H2($K$) = 1.034, values consistent with a low gravity.  We consider this object a high probability member of $\beta$ Pic.  We use its photometric distance estimate, an age of 24 $\pm$ 3 Myr \citep{bell15}, the spectral type-$K_{\rm S}$ bolometric correction relation for young objects from \cite{fil15}, and the evolutionary models of \cite{sau08} to find a mass range of 9$-$12 $M_{\rm Jup}$, placing WISEA J235422.31$-$081129.7 in the planetary mass regime if a true $\beta$ Pic member.  \cite{allers13} did not find suitable low and intermediate gravity standards for spectral type L5.  We thus compare the near-infrared spectrum of WISEA J235422.31$-$081129.7 to the low and intermediate surface gravity L6 standards in Figure 7.   

\begin{figure*}
\plotone{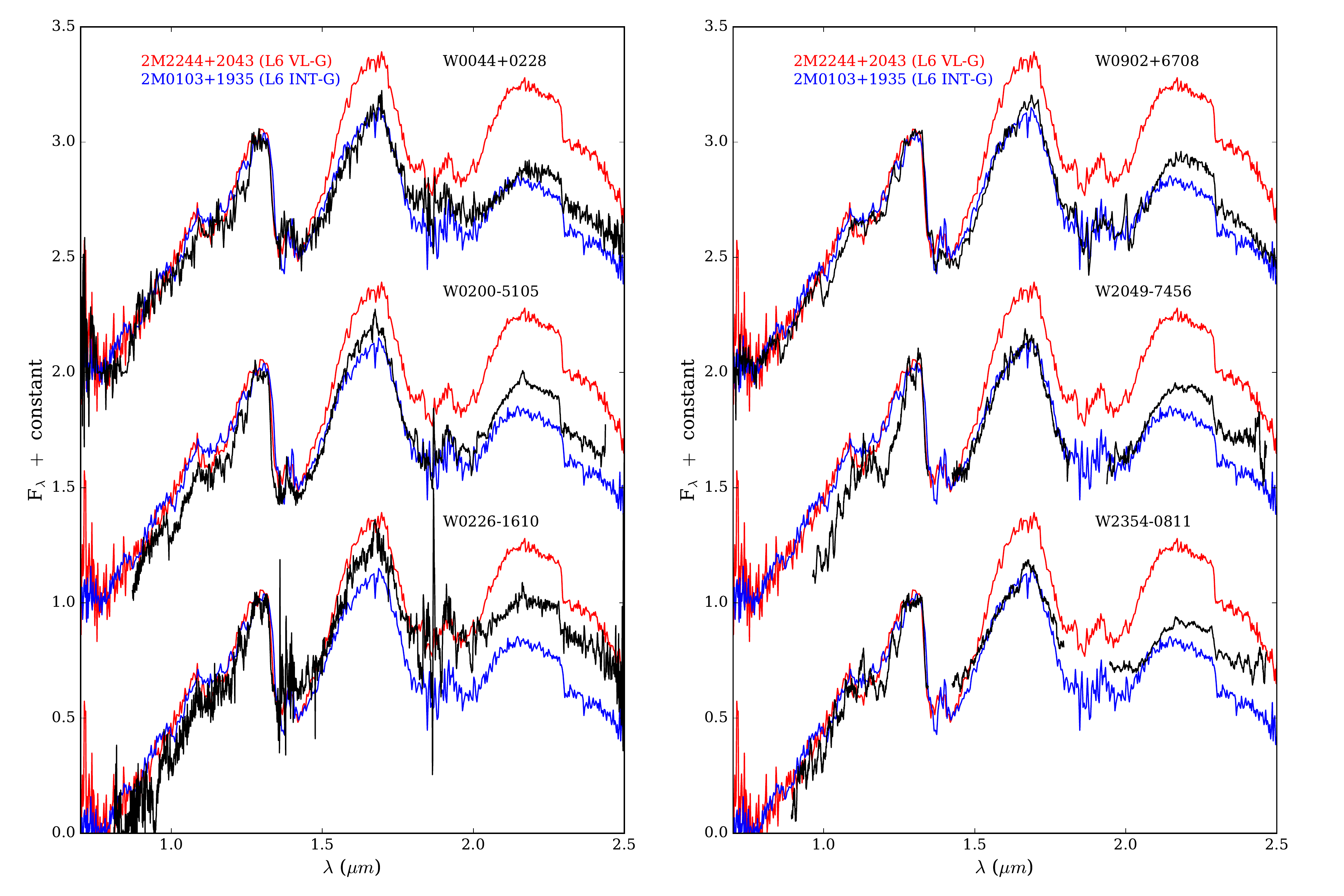}
\caption{Near-infrared spectra of several newly proposed moving group members discovered with this survey compared to the low gravity standards suggested in \cite{allers13}.  All spectra are normalized between 1.27 and 1.29 $\mu$m.  The standards used for comparison are 2MASSI J0103320$+$193536 (L6 INT-G; \citealt{cruz04}) and 2MASSW J2244316$+$204343 (L6 VL-G; \citealt{loop08}).}  
\end{figure*}

\section{Conclusions}

We have used the unique near- and mid-infrared colors of young, late-type L dwarfs to identify 50 new late-type L dwarf candidates, 47 of which we have confirmed spectroscopically as L dwarfs.  We assert that two objects (WISEA J004403.39$+$022810.6 and WISEA J235422.31$-$081129.7) are likely $\beta$ Pic members based on their membership probabilities from BANYAN II and the convergent point tool of \cite{rod13}, youthful spectroscopic characteristics, and distance estimates.  If true $\beta$ Pic members, we estimate that both of these objects have masses in the planetary mass regime.  We also find three highly likely members of ABDor (WISEA J020047.29$-$510521.4, WISEA J022609.16$-$161000.4, and WISEA J090258.99$+$670833.1), one medium to high probability member of Argus (WISEA J072352.62$-$330943.5), and one moderate-probability member of TucHor (WISEA J204902.80$-$745613.5).  We have also identified two potential late-L type Hyades members (WISEA J043642.75$+$190134.8 and WISEA J044105.56$+$213001.5).  These objects, if confirmed would be some of the lowest mass members of these groups. Because brown dwarfs cool as they age, they do not obey a simple mass-luminosity relationship like stars.  Instead, brown dwarfs follow mass-luminosity-age relation, making age a vital parameter for the determination of brown dwarf physical properties. This sample of newly discovered potential moving group and cluster members thus provides indispensable benchmarks for investigating the evolution of low-mass objects and constraining low-mass evolutionary models.  

The limiting factor in this search was the depth of the 2MASS catalog.  Expanding this search to include deeper near-infrared catalogs, such as UKIDSS \citep{law07} or the VISTA Hemisphere Survey (VHS; PI McMahon, Cambridge, UK) would undoubtedly reveal more late-L type members of the Solar neighborhood.   

\acknowledgments

We wish to thank our anonymous referee for their helpful comments and suggestions that improved the quality of this paper.  The authors would like to thank Katelyn Allers for help with the ARCoIRIS observations and the reduction of ARCoIRIS and FIRE data, and Scott Sheppard for useful discussions regarding FIRE observations.  This paper includes data gathered with the 6.5 meter Magellan Telescopes located at Las Campanas Observatory, Chile and is based in part on observations at Cerro Tololo Inter-American Observatory, National Optical Astronomy Observatory (NOAO Prop. ID: 2016B-0003; PI: A.\ Schneider), which is operated by the Association of Universities for Research in Astronomy (AURA) under a cooperative agreement with the National Science Foundation. This publication makes use of data products from the {\it Wide-field Infrared Survey Explorer}, which is a joint project of the University of California, Los Angeles, and the Jet Propulsion Laboratory/California Institute of Technology.  {\it WISE} is funded by the National Aeronautics and Space Administration.  This publication makes use of data products from the Two Micron All Sky Survey, which is a joint project of the University of Massachusetts and the Infrared Processing and Analysis Center/California Institute of Technology, funded by the National Aeronautics and Space Administration and the National Science Foundation.  This research has made use of the SIMBAD database, operated at CDS, Strasbourg, France.  This research has benefitted from the M, L, T, and Y dwarf compendium housed at dwarfarchives.org.  The authors wish to recognize and acknowledge the very significant cultural role and reverence that the summit of Mauna Kea has always had within the indigenous Hawaiian community.  We are most fortunate to have the opportunity to conduct observations from this mountain.

\floattable
\rotate
\begin{deluxetable*}{lcccccccccccc}
\tablecaption{Moving Group Membership Summary}
\tablehead{
\colhead{AllWISE Designation} & \colhead{BANYAN II} &\colhead{$d_{\rm pred}$\tablenotemark{a}} & \colhead{LACEwING} & \colhead{$d_{\rm pred}$\tablenotemark{a}} & \colhead{CP} & \colhead{$d_{\rm pred}$\tablenotemark{a}} \\
 & (\%) & (pc) & (\%) & (pc) & (\%) & (pc) }
\startdata
J004403.39$+$022810.6 & $\beta$ Pic (78) & 33 $\pm$ 4 & \dots & \dots & Columba (99), $\beta$ Pic (97) & 46, 36 \\
J005811.69$-$565332.1 & $\beta$ Pic (27), Argus (31) & 19 $\pm$ 2, 25 $\pm$ 3 & TucHor (48), ABDor (40) & 29 $\pm$ 3, 29 $\pm$ 2 & Car-Near (50) & 33 \\
J020047.29$-$510521.4 & ABDor (98) & 22 $\pm$ 2 & TucHor (28), ABDor (53) & 24 $\pm$ 3, 23 $\pm$ 2 & ABDor (54) & 23 \\
J022609.16$-$161000.4 & $\beta$ Pic (12), ABDor (85) & 23 $\pm$ 3, 36 $\pm$ 3 & TucHor (36), ABDor (36) & 34 $\pm$ 3, 37 $\pm$ 1 & ABDor (99) & 36\\
J023749.81$-$260543.8 & \dots & \dots & \dots & \dots & \dots & \dots \\
J041232.77$+$104408.3 &  \dots & \dots & Hyades (87) & 49 $\pm$ 13 & \dots & \dots \\
J043642.75$+$190134.8 &  \dots & \dots & Hyades (87) & 47 $\pm$ 5 & Car-Near (55) & 37 \\
J043718.77$-$550944.0 &  $\beta$ Pic (94) & 17 $\pm$ 3 & TucHor (20), ABDor (20) & 32 $\pm$ 9, 18 $\pm$ 6 & $\beta$ Pic (93), TWA (88) & 19, 22 \\
J044105.56$+$213001.5 &  Argus (14) & 22 $\pm$ 5 & Hyades (100) & 46 $\pm$ 6 & Car-Near (74) & 36 \\
J045900.42$-$285338.3 & $\beta$ Pic (29), Argus (55) & 9 $\pm$ 3, 19 $\pm$ 4 & \dots & \dots & Car-Near (97) & 28 \\
J072352.62$-$330943.5 & Argus (53) & 26 $\pm$ 4 & Argus (25) & 31 $\pm$ 2 & Car-Near (97) & 36 \\
J081322.19$-$152203.2 & Argus (87) & 15 $\pm$ 2  & \dots & \dots & Car-Near (97) & 17 \\
J090258.99$+$670833.1 & ABDor (56) & 23 $\pm$ 2 & \dots & \dots & Columba (83) & 23 \\
J130523.06$-$395104.9 & Argus (95) & 21 $\pm$ 2  & \dots & \dots & TucHor (80), Car-Near (59) & 18, 27\\
J131845.58$+$362614.0 & \dots & \dots & \dots & \dots & Car-Near (90) & 72 \\
J145642.68$+$645009.7 & ABDor (52) & 16 $\pm$ 2 & \dots & \dots & ABDor (71), Columba (65) & 17, 21 \\
J162341.27$-$740230.4 & $\beta$ Pic (58), ABDor (36) & 11 $\pm$ 1, 14 $\pm$ 1 & Argus (32), ABDor (38) & 13 $\pm$ 1, 14 $\pm$ 2 & ABDor (70) & 14 \\
J173453.90$-$481357.9 & Argus (45) & 13 $\pm$ 2 & Argus (36), ABDor (24) & 17 $\pm$ 2, 28 $\pm$ 2 & Car-Near (63) & 20 \\
J204902.80$-$745613.5 & $\beta$ Pic (29), TucHor (42) & 29 $\pm$ 4, 49 $\pm$ 5 & TucHor (31), ABDor (22) & 37 $\pm$ 7, 42 $\pm$ 3 & Columba (98), TucHor (81) & 42, 37 \\
J235422.31$-$081129.7 & $\beta$ Pic (94) & 29 $\pm$ 3 & ABDor (23) & 47.5 $\pm$ 1 & $\beta$ Pic (100), TWA (96) & 30, 31\\
\enddata
\tablenotetext{a}{Predicted distance if indeed a member of the moving group (or groups) listed in the ``BANYAN II'', ``LACEwING'', or ``CP'' columns.  If more than one value is listed, their order corresponds to the order of groups listed in the previous column.}
\end{deluxetable*}

\end{document}